\documentclass[preprintnumbers,nofootinbib,showkeys,showpacs,amsmath,amssymb,aps]{revtex4}
\usepackage{amsmath,amssymb,mathrsfs,graphics,epsfig,subfigure}
\usepackage{color}
\usepackage{float}
\usepackage{multirow}
\usepackage{booktabs}

\usepackage{enumitem}
\usepackage{hyperref}
\hypersetup{colorlinks=true,linkcolor=blue,citecolor=magenta}
\begin{document}
	\newcommand {\nn} {\nonumber}
	\renewcommand{\baselinestretch}{1.3}

\title{Precession of spherical orbits for the spacetime without $\mathbb{Z}_2$ symmetry induced by NUT charge}
	
	\author{Xiang-Cheng Meng, Shan-Ping Wu, Shao-Wen Wei \footnote{Corresponding author. E-mail: weishw@lzu.edu.cn}}

	\affiliation{Key Laboratory of Quantum Theory and Applications of MoE, Lanzhou Center for Theoretical Physics,\\
	 Key Laboratory of Theoretical Physics of Gansu Province, Gansu Provincial Research Center for Basic Disciplines of Quantum Physics, Lanzhou University, Lanzhou 730000, China\\
		Institute of Theoretical Physics \& Research Center of Gravitation, Lanzhou University, Lanzhou 730000, China}

\begin{abstract}
Astrophysical evidence has hinted at the existence of a nonzero NUT charge, which breaks the $\mathbb{Z}_2$ symmetry of spacetime and induces novel features in geodesics. In this work, we investigate the Lense-Thirring precession of the spherical orbits in the Kerr-Taub-NUT spacetime, with particular emphasis on its connection to recent observations of black hole jet precession. We analyze the reflection symmetry breaking in trajectories of the spherical orbits and extract their precession angular velocity. It is worth noting that in the absence of spin, the spherical orbits reduce to tilted circular orbits without precession, whereas for nonzero spin, the precession angular velocity increases with the absolute value of the NUT charge. We then model the motion of particles near the warp radius of a tilted accretion disk using the spherical orbits and constrain the black hole parameter space based on the observed jet precession of M87*. The results indicate that regions with low spin and large NUT charge were excluded, and that the jet precession measurements cannot distinguish the sign of the NUT charge. The excluded region is larger for retrograde accretion disks than for prograde ones. We also find that this observation do not allow a clear distinction between black holes and naked singularities. Moreover, we also explore how black hole parameters influence the structure of accretion disk. These results have important theoretical and astronomical significance for us to deeply understand NUT space-time.
\end{abstract}

	
\maketitle

\tableofcontents

\section{Introduction}\label{secIntroduction}
    Taub-NUT solution, which is an exact solution to the vacuum Einstein equations, exhibits intriguing properties. Taub was the first to derive a solution that covers a portion of the full spacetime~\cite{TaubAH}, which can be interpreted as a homogeneous but anisotropic cosmological model~\cite{Wheeler1962,Brill1964}. Subsequently, Newman, Tamburino, and Unti obtained the full Taub-NUT solution as a generalization of the Schwarzschild metric~\cite{TaubNUT}. While the Schwarzschild metric involves a single parameter $M$ interpreted as mass, the NUT metric introduces an additional parameter $n$, commonly referred to as the NUT charge. The presence of this charge renders the spacetime highly unusual, to the extent that Misner famously described it as a counter-example to almost anything \cite{Misner1967}.

    One of the most troubling aspects of the Taub-NUT spacetime is the existence of a string-like singularity along the axis. Two radically different approaches have been proposed to deal with it, both with unsatisfactory consequences. Misner constructed a globally regular version of the NUT spacetime~\cite{Misner1963}, in which the singularity is removed entirely at the cost of imposing periodicity on the time coordinate. However, this results in the existence of closed timelike curves, thus violating the causality. To avoid such causality issues, Bonnor instead retained the singularity at $\theta = \pi$ and interpreted it as semi-infinite massless rotating rod~\cite{Bonnor1969}. In this interpretation, the closed timelike curves are confined to a small neighborhood of the singularity. A more physically motivated approach involves restricting the coordinate range to exclude the causally pathological regions containing the closed timelike curves. The resulting spacetime manifold contains holes, and the geodesics become incomplete, in the sense that not all of them can be extended to arbitrary affine parameters.

    Despite its imperfections, Taub-NUT spacetime may still serve as a useful model of certain astrophysical scenarios. Previous study has shown that exact solutions to the vacuum Einstein equations often carry deep physical significance. For instance, the Schwarzschild solution was initially considered as unphysical due to the event horizon at $r = 2M$, but observational evidence now strongly supports the existence of black holes~\cite{LIGO,Akiyama1,Akiyama2}, and the Schwarzschild metric remains an excellent approximation of spacetime geometry in the vicinity of compact objects (though in practice, its rotating extension, the Kerr metric, is more commonly used). Motivated by this viewpoint, it is reasonable to explore the Taub-NUT metric as a potential model for the spacetime around certain astrophysical sources. A natural way to probe the spacetime geometry is through its geodesics. In particular, for timelike geodesics, the Lense-Thirring (LT) precession provides an important observational signature \cite{Thirring1918,LenseThirring1918,Ciufolini2004,Iorio2025}. Unfortunately, the spherical orbits in the Taub-NUT spacetime that we intend to study, which are closely related to astrophysical precession, do not exhibit LT precession~\cite{VKagramanova}. In contrast, this effect is well-established in the Kerr spacetime~\cite{Wilkins,Goldstein,Dymnikova,Shakura,ETeo,PRana,Kopek}. Thus, to study the influence of the NUT charge on LT precession, one must turn to the rotating generalization of the Taub-NUT solution, namely, the Kerr-Taub-NUT (KTN) spacetime.

    KTN solution, as a special case of the Pleba\'nski-Demia\'nsk solution family belonging to Petrov type D~\cite{Plebanski1976}, usually describes a rotating black hole endowed with NUT charge~\cite{Miller1973}. Its geodesic structure has been widely investigated, including null geodesics related to black hole shadows~\cite{Abdujabbarov2013,Zhang2021} and topology of light rings~\cite{Wu2023}. Timelike geodesics, particularly circular orbits, have also been analyzed~\cite{Chakraborty20141307,Pradhan2015,Cebeci2019,Chakraborty2019}, with some analytic solutions provided~\cite{VKagramanova,Hackmann2009,Cebeci2016}. A particularly noteworthy feature of the KTN spacetime is that the presence of the NUT charge breaks the $\mathbb{Z}_2$ symmetry of the spacetime, leading to fundamental differences in its geodesic structure compared to the Kerr case. In Kerr spacetime, circular orbits naturally lie on the equatorial plane, and spherical orbits, which have a constant radial coordinate, are tilted with respect to these equatorial circular orbits \cite{Zahrani}.  Their conserved quantities and trajectories can be directly obtained from the equations governing the spherical orbits. However, in the KTN spacetime, due to the influence of the NUT charge, circular orbits are no longer confined to the equatorial plane. Instead, they are located along planes parallel to the equatorial plane. As a result, in order to study the spherical orbits in this spacetime, one must first determine the positions of these non-equatorial circular orbits. In addition, these orbits no longer respect reflection symmetry about the circular orbit in the $\theta$ direction, which reflects the breaking of the underlying $\mathbb{Z}_2$ symmetry of the spacetime. It is thus evident that the spherical orbits in the KTN spacetime possess unique kinematic features. Building on this, we investigate the spherical orbits and their associated LT precession. Although this effect has been discussed in several previous studies~\cite{Chakraborty2014,Chakraborty2015}, our study goes beyond purely theoretical analysis. We aim to use these theoretical orbital motions to model particle dynamics relevant to astrophysical systems, and combine them with observational data to explore the possible existence of the NUT charge or place constraints on it.

    From the perspective of astrophysical observations, the study of the KTN black hole is of considerable significance. The NUT charge, also referred to as the gravitomagnetic charge, can be regarded as a gravitational analogue of the Dirac magnetic monopole~\cite{Dirac1931}, and its physical existence remains an intriguing possibility. For example, in the case of the X-ray binary system GRO J1655-40, one study inferred the spin and other parameters using three independent methods under the assumption of a nonzero gravitomagnetic monopole~\cite{Chakraborty2018}. The results were mutually consistent and yielded a black hole mass compatible with independent measurements. In contrast, when the monopole disappears, the three methods produced significantly different spin values for the accreting black hole. This strongly suggests the possible presence of a nonvanishing gravitomagnetic monopole in this system. Subsequent research explained the jet power and radiative efficiency of black holes in several X-ray binary systems, including GRO J1655-40, within the framework of the KTN spacetime~\cite{Narzilloev2023}. These results indicated that the presence of a gravitomagnetic monopole could offer a plausible explanation for these observational characteristics. Interestingly, a warped accretion disk has also been observed in the GRO J1655-40 system~\cite{Martin2008}. To interpret this feature, recent studies have considered tilted thin accretion disks around KTN black holes and explored the influence of the NUT charge on the disk's tilt angle~\cite{Sen2024}. Such tilted structures may significantly affect X-ray spectral and timing features via the LT effect, offering potential observational diagnostics.

    Furthermore, the observation of the M87 black hole shadow by the Event Horizon Telescope (EHT) has been employed to test the possible existence of a gravitomagnetic monopole~\cite{Ghasemi2021}. The results suggest that the presence of such a monopole is compatible with the EHT data, and therefore, the possibility that the central black hole in M87 carries a nonzero NUT charge cannot be ruled out. Recently, based on 22 years of radio observations, Cui et al. reported a precession period of approximately 11 years for the M87 jet, strongly indicating the presence of a supermassive rotating black hole with a tilted accretion disk at the galactic center~\cite{Cui}. Tilted accretion disks have been observed in various systems, including protostars, X-ray binaries, and active galactic nuclei (AGN)~\cite{Papaloizou,Herrnstein,Begelman,Wijers,Chiang,Martin2008,Lodato,Casassus,Putten}. Investigating such structures helps to uncover the physical origins of astrophysical phenomena such as black hole jet precession. A more detailed discussion about the tilted disks is presented in Sec. \ref{secDisk}. Although it is challenging to construct precise models of tilted accretion disks around rotating black holes, simplified theoretical models can still capture the essential physics. In the Kerr spacetime background, we modeled the motion of particles in tilted disks using the spherical orbits and used the observed precession period of the jet to constrain the black hole spin parameter~\cite{Wei}. Given the strong observational support for the existence of a gravitomagnetic monopole in certain sources, we propose using the more general KTN geometry to study the jet precession of M87, thereby broadening the scope of the investigation. In this work, we model the structure of a tilted thin accretion disk based on the spherical orbits in the KTN spacetime, constrain the black hole parameters using the observed jet precession, and explore how these parameters affect the structure of disk.

    Since the analysis of the spherical orbits requires prior knowledge of the circular orbits' position, our paper is organized as follows. In Sec.~\ref{secKNTCO}, we analyze circular orbits in the KTN spacetime. In Sec.~\ref{secPOSO}, we study the LT precession of the spherical orbits, which exhibit broken symmetry in the $\theta$ direction. In Section \ref{secDisk}, we use the spherical orbits to model motion of particle in a tilted accretion disk around the central black hole of M87, constrain black hole parameters with jet precession observations, and explore their impact on disk structure. Section \ref{secConclusions} presents a summary and outlook. we adopt the metric convention $(-,+,+,+)$ and use geometrical units with $G=c=1$ in addition to recovering dimensionality in Sec. \ref{secDisk}.
	
\section{Non-equatorial circular orbits in the Kerr-Taub-NUT spacetime}\label{secKNTCO}

In general, circular orbits in the equatorial plane serve as a foundation for exploring other more general spherical orbits. However, due to the presence of the NUT charge, the $\mathbb{Z}_2$ symmetry is broken, which leads to that circular orbits lie on non-equatorial planes. In this section, we would like to examine the properties of the circular orbits in the KTN spacetime. Specifically, we determine the position, energy, and angular momentum of these non-equatorial circular orbits, as well as the innermost stable circular orbits (ISCOs), which play a central role in the subsequent analysis of accretion disk structures.

The KTN spacetime is a stationary and axisymmetric solution to the vacuum Einstein field equations, expressed in Boyer-Lindquist coordinates as follows \cite{Demianski,Griffiths},
\begin{equation}
	ds^2 = -\frac{\Delta} {\Sigma} (dt - {\chi}d {\phi})^2+\frac{\Sigma}{\Delta}dr^2+\Sigma d{\theta}^2+\frac{ \sin^2{\theta}}{\Sigma}\left(a\,dt-(r^2+n^2+a^2)d{\phi}\right)^2,
	\label{eqmetric}
\end{equation}
where the metric functions read
\begin{align*}
	\Sigma &= r^2+(n+a \cos \theta)^2,\\
	\Delta &= r^2-2M r +a^2-n^2,\\
	\chi &= a \sin^2 \theta - 2 n \cos \theta.
\end{align*}
Besides the black hole's mass $M$ and spin $a$, the solution includes the NUT charge $n$, which is also referred to as the magnetic mass or the gravitomagnetic monopole. When the spin $a$ vanishes, the metric reduces to a vacuum solution to Einstein's equations that extends the Schwarzschild metric \cite{TaubAH,TaubNUT}. The presence of the NUT charge renders the spacetime no longer globally asymptotically flat, and introduces string singularity on the symmetric axis. Furthermore, it breaks $\mathbb{Z}_2$ symmetry of spacetime; the metric is no longer invariant under the transformation $\theta \rightarrow \pi - \theta$. As a result, geodesics, which are sensitive to the spacetime geometry, undergo significant changes. Our aim is to examine how the NUT charge influences geodesics that admit the spherical orbits and to identify potential imprints of this effect in astronomical observations.

\begin{figure}[!htbp]
	\centering{
		\includegraphics[width=10cm]{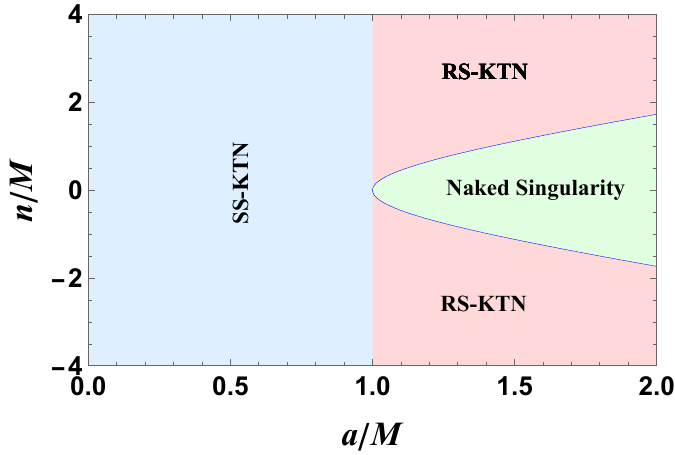}
		\caption{Parameter regions for different solutions. The blue region ($a<M$) corresponds to the scenario of slowly spinning KTN (SS-KTN) black holes. The red region ($a>M$) represents the case of rapidly spinning KTN (RS-KTN) black holes with two horizons. The green region corresponds to the naked singularity scenario without horizon. The blue thin line serves as the boundary between the red and green regions, representing the extreme KTN black hole case with a degenerate horizon.
		}\label{figHorizon1}
	}
\end{figure}

The event horizons of the KTN black hole are located at $r = M\pm \sqrt{M^2+n^2-a^2}$. The naked singularities correspond to $M^2 + n^2 - a^2 < 0$, where the spacetime has no horizon. The ring singularities exist at $r = 0$ and $\theta = \arccos(-n/a)$ when the condition $n^2 \leq a^2$ is satisfied. It is worth noting that when the NUT charge is non-zero, the event horizon can exist even for spin $a > M$, quite different from the Kerr black hole cases. We refer to this as a rapidly spinning KTN black holes. In contrast, the case $a \leq M$ corresponds to slowly spinning KTN black holes. In Fig. \ref{figHorizon1}, we clearly present the parameter regions for the slowly spinning KTN black holes, rapidly spinning KTN black holes, and naked singularities.

Circular orbits are a special class of the spherical orbits with zero tilt angle, whose equations of motion follow the general spherical case with an additional constraint in the $\theta$ direction. In this section and Sec. \ref{secPOSO}, we will derive the equations of motion for circular and spherical orbits, respectively. These first-order equations simplify the subsequent analysis and allow for analytic expressions of conserved quantities such as energy and angular momentum \cite{XiangChengMeng2025ImprintsOB}.

For a massive test particle, the Hamiltonian takes the form
\begin{equation}\label{eqHamiltonian}
	H=\frac{1}{2}g^{\mu\nu}p_{\mu}p_{\nu},
\end{equation}
and it satisfies the momentum normalization condition
\begin{equation}\label{eqpsquar1}
	p^{\mu}p_{\mu}=-1.
\end{equation}
Substituting the Hamiltonian Eq. \eqref{eqHamiltonian} into the canonical equations
\begin{equation}\label{eqregulareq}
	\dot{x}^{\mu} = \frac{\partial H}{\partial p_\mu}, \ \dot{p}_{\mu} = -\frac{\partial H}{\partial x^\mu},
\end{equation}
we obtain the four-velocity $\dot{x}^{\mu}$ and momenta $p_{\mu}$ as
\begin{align}
	\label{eqtdot} \dot{t} &= g^{tt}p_{t} + g^{t\phi}p_{\phi},\\
	\label{eqphidot} \dot{\phi} &= g^{t\phi}p_{t} + g^{\phi \phi}p_{\phi},\\
	\label{eqrdot} \dot{r} &= g^{rr}p_{r},\\
	\label{eqthetadot} \dot{\theta}&= g^{\theta \theta}p_{\theta}, \\
\end{align}
and
\begin{align}
	\label{eqptdot}\dot{p}_{t} = 0,\hspace{5cm}\kern 0.4em\\
	\label{eqpphidot} \dot{p}_{\phi} = 0,\hspace{5cm}\kern 0.4em\\
	\label{eqprdot} \dot{p}_{r}  = -\frac{1}{2}\left(p_{t}^{2} \partial_{r}g^{tt} +2p_{t}p_{\phi} \partial_{r}g^{t\phi}+p_{\phi}^{2} \partial_{r}g^{\phi \phi} +p_{r}^{2} \partial_{r}g^{rr} +p_{\theta}^{2} \partial_{r}g^{\theta \theta}\right),\\
	\label{eqpthetadot} \dot{p}_{\theta} = -\frac{1}{2}\left(p_{t}^{2} \partial_{\theta }g^{tt}+2p_{t}p_{\phi}\partial_{\theta}g^{t\phi} + p_{\phi}^{2} \partial_{\theta}g^{\phi \phi} +p_{r}^{2} \partial_{\theta}g^{rr} +p_{\theta}^{2} \partial_{\theta}g^{\theta \theta} \right),
\end{align}
From Eqs. \eqref{eqptdot} and \eqref{eqpphidot}, we can directly obtain that energy and angular momentum are conserved quantities, expressed as
\begin{align}
	\label{eqdefE} p_{t}&=-E,\\
	\label{eqdefL} p_{\phi}&=L,
\end{align}
which are associated with the Killing vectors $\partial_t$ and $\partial_\phi$, respectively. Combining with above equations, it yields
\begin{equation}\label{eqtimelike}
	g^{rr}p_r^2+g^{\theta \theta}p_{\theta}^2=\mathcal{V}(r,\theta),
\end{equation}
where the effective potential $\mathcal{V}(r,\theta)$ is defined as
\begin{equation}
	\mathcal{V}(r,\theta)=-g^{tt}E^2+2g^{t\phi}EL-g^{\phi \phi}L^2-1.
\end{equation}
Since both circular and spherical orbits lie on a spherical surface with constant radius $r = \text{const}$, i.e., $\dot{r}=\ddot{r}=0$, it leads to the conditions
\begin{equation}\label{eqrconstraint}
	p_{r} =\dot{p}_{r}= 0.
\end{equation}
Note that the above calculations are valid for both types of orbits. In Sec. \ref{secPOSO}, we will directly use these results for the spherical orbits without repeating the process.

In Kerr-like spacetimes, circular orbits lie in the equator with radial coordinate $r=\text{const}$. Further study of the spherical orbits only requires knowing their tilt angle relative to the equatorial circular orbits. However, when the NUT charge is nonzero, circular orbits lie on the non-equatorial planes~\cite{Jefremov,Mukherjee}.  Here, we denote the angular position of the circular orbit to be determined as $\theta_0$. The $\theta$-motion of circular orbit is thus fixed at $\theta = \theta_0$, satisfying the constraint $\dot{\theta}=\ddot{\theta}=0$, i.e.,
\begin{equation}\label{eqcircularthetacon}
    p_{\theta}|_{\theta=\theta_{0}}=\dot{p_{\theta}}|_{\theta=\theta_{0}}=0.
\end{equation}
By imposing the constraint Eqs. \eqref{eqrconstraint} and \eqref{eqcircularthetacon}, we find that the energy $E$ and angular momentum $L$ of the circular orbits must satisfy
\begin{align}
    \label{eqpthetadotCircular} \left.\left(E^{2} \partial_{r}g^{tt} -2EL \partial_{r}g^{t\phi}+L^{2} \partial_{r}g^{\phi \phi}\right)\right|_{\theta=\theta_{0}}&=0,\\
    \label{eqtimelikeCircular}\left.\left(-g^{tt}E^2+2g^{t\phi}EL-g^{\phi \phi}L^2-1\right)\right|_{\theta=\theta_{0}}&=0.
\end{align}
Solving these two equations yields analytic expressions for the energy and angular momentum
\begin{align}
	\label{eqHECirc}E=\left. \frac{1}{\sqrt{-g^{tt}-2g^{t\phi}\chi -g^{\phi \phi}\chi ^2}} \right|_{\theta =\theta_{0}},\\
	\label{eqHLCirc}L=\left. \frac{-\chi}{\sqrt{-g^{tt}-2g^{t\phi}\chi -g^{\phi \phi}\chi ^2}} \right|_{\theta =\theta_{0}},
\end{align}
where
\begin{equation}
	\chi =\frac{p_{\phi}}{p_t}=\left. \frac{-\partial _rg^{t\phi}\pm \sqrt{\left( \partial _rg^{t\phi} \right) ^2-\partial _rg^{tt}\partial _rg^{\phi \phi}}}{\partial _rg^{\phi \phi}} \right|_{\theta =\theta_{0}},
\end{equation}
It should be emphasized that the ``$\pm$'' signs in the expressions for $E$ and $L$ correspond to prograde and retrograde orbits, respectively. We take the direction of the black hole's spin as the reference, and thus the prograde orbits have positive angular momentum, while retrograde orbits have negative angular momentum.

To determine the angular position $\theta_0$ of the circular orbit, we insert the constraint Eq. \eqref{eqcircularthetacon} into Eq.~\eqref{eqpthetadot}, yielding the condition
\begin{equation}\label{eqtheta0}
    \left.\left(E^{2} \partial_{\theta }g^{tt}-2EL\partial_{\theta}g^{t\phi} + L^{2} \partial_{\theta}g^{\phi \phi} \right)\right|_{\theta=\theta_{0}}=0.
\end{equation}
Since $E$ and $L$ are already determined, this becomes an equation relating the orbit radius $r$ and position $\theta_0$. For a given $r$, the value of $\theta_0$ can be obtained by solving this equation.

In summary, once the circular orbit radius $r$, the spin $a$, and the NUT charge $n$ of black hole are specified, the energy $E$, angular momentum $L$, and the angular position $\theta_{0}$ of the orbit are fully determined. The motion in the $\phi$ direction can then be obtained by solving the following equation of motion
\begin{align}
    \label{eqCirctdot} \dot{t} &= -g^{tt}E + g^{t\phi}L,\\
	\label{eqCircphidot} \dot{\phi} &=-g^{t\phi}E + g^{\phi \phi}L.
\end{align}

\subsection{Energy and angular momentum}\label{subseccircposi}

It is evident from the derivation that for a given orbital radius, determining the orbital position, energy, and angular momentum is essential for solving the trajectory. Accordingly, we computed the variation of circular orbits' angular position $\theta_0$, energy $E$, and angular momentum $L$ with respect to the orbital radius, as given by Eqs. \eqref{eqtheta0}, \eqref{eqHECirc} and \eqref{eqHLCirc}. Note that the inverse metric in Eq. \eqref{eqmetric} is used in the calculations.

\begin{figure}[!htbp]
    \centering{
			\subfigure[]{\includegraphics[width=5.8cm]{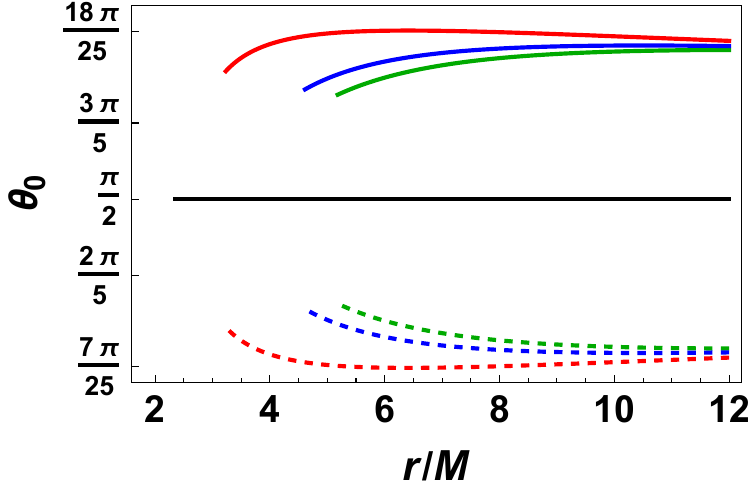}\label{figpCircOrbitTheta2a}}
			\subfigure[]{\includegraphics[width=5.9cm]{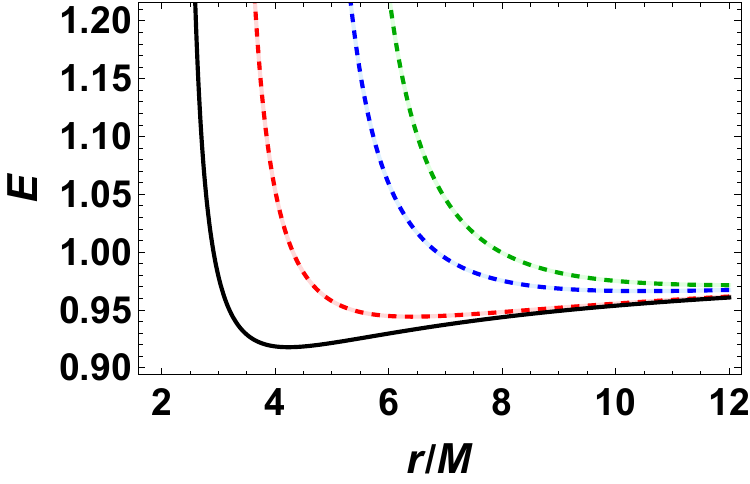}\label{figpCircOrbitE2b}}
			\subfigure[]{\includegraphics[width=5.7cm]{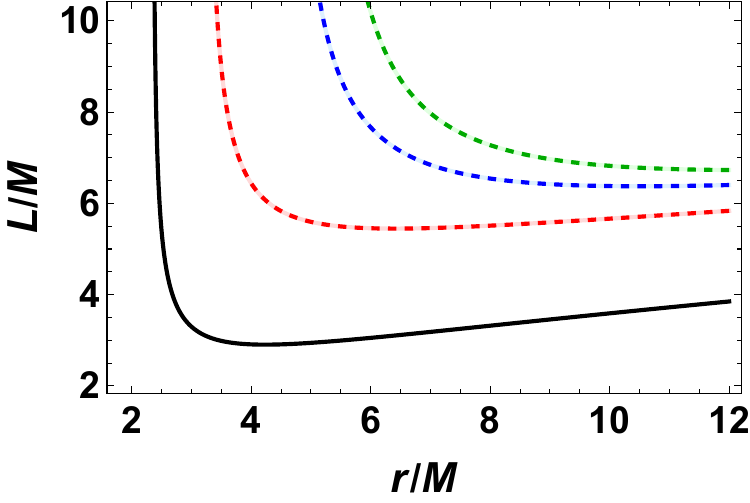}\label{figpCircOrbitL2c}}\\
			\subfigure[]{\includegraphics[width=5.8cm]{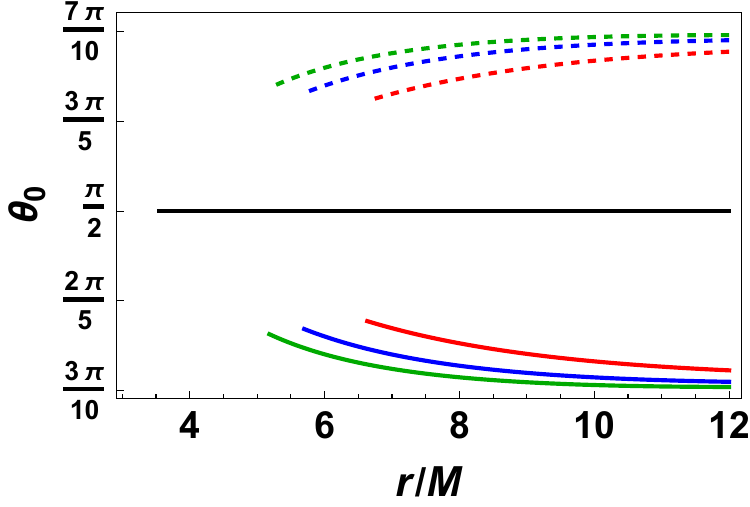}\label{fignCircOrbitTheta2d}}
            \subfigure[]{\includegraphics[width=5.9cm]{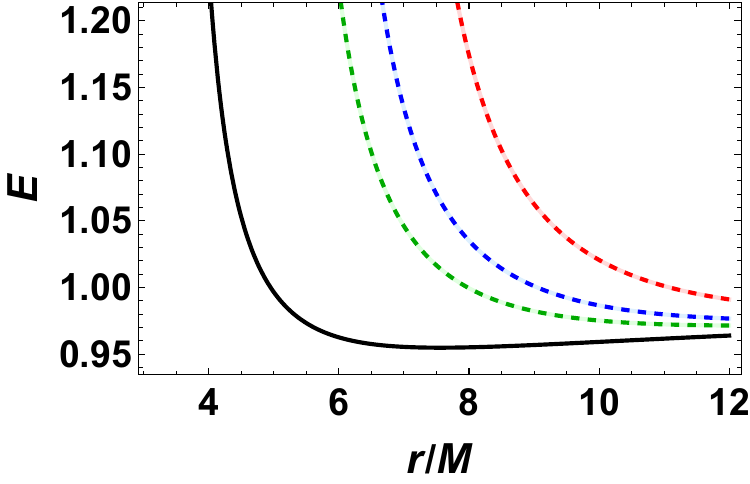}\label{fignCircOrbitE2e}}
            \subfigure[]{\includegraphics[width=5.9cm]{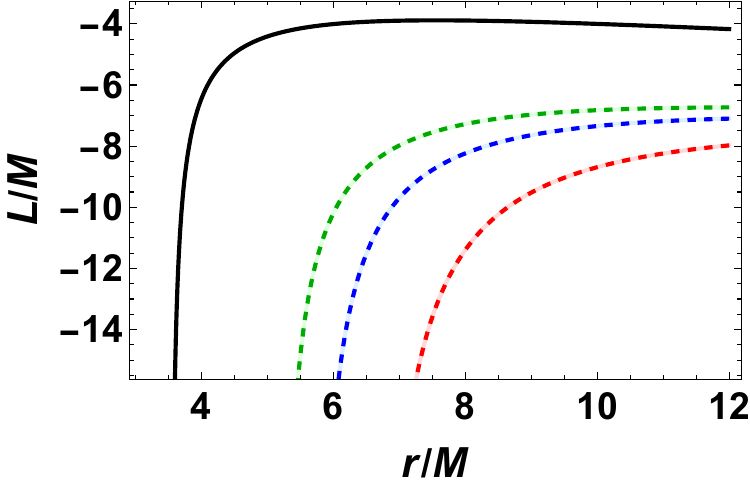}\label{fignCircOrbitL2f}}
			\caption{The angular position $\theta_0$, energy $E$, and angular momentum $L$ of circular orbits vary with the radius $r$ in the Kerr, Taub-NUT, SS-KTN, and RS-KTN spacetimes. The upper and lower rows correspond to prograde and retrograde circular orbits. Black solid curve represents the Kerr case with $a/M = 0.5$. The (light) green, (light) blue, and (light) red curves correspond to Taub-NUT, SS-KTN, and RS-KTN spacetimes with $a/M = 0$, $0.5$, and $1.5$, respectively. The dashed curves for $n/M =-2$ and solid curves for $n/M =2$. (a) $\theta_{0}$ vs $r$ for prograde orbits. (b) $E$ vs $r$ for prograde orbits. (c) $L$ vs $r$ for prograde orbits. (d) $\theta_{0}$ vs $r$ for retrograde orbits. (e) $E$ vs $r$ for retrograde orbits. (f) $L$ vs $r$ for retrograde orbits.}\label{figCOle2}
    }
\end{figure}

In Fig.~\ref{figCOle2}, we illustrate the behaviors of $\theta_{0}, E$ and $L$ in the Kerr, Taub-NUT, SS-KTN, and RS-KTN spacetimes. We observe that the presence of the NUT charge causes the angular position $\theta_0$ to deviate from the equatorial plane. Specifically, for prograde orbits, $\theta_0 > \pi/2$ when $n > 0$, and $\theta_0 < \pi/2$ when $n < 0$. For retrograde orbits, the trend is reversed. We thus summarize this behavior as follows: when $nL > 0$, the circular orbit lies in the southern hemisphere ($\theta_0 > \pi/2$); when $nL < 0$, it lies in the northern hemisphere ($\theta_0 < \pi/2$). Notably, the circular orbits corresponding to positive and negative NUT charges are always symmetric with respect to the equatorial plane. Moreover, the greater the absolute value of the NUT charge $n$, the larger the deviation of the circular orbit from the equatorial plane. The black hole spin $a$ also influences the position $\theta_0$: for prograde orbits, increasing $a$ enhances the deviation of $\theta_0$ from $\pi/2$, whereas for retrograde orbits, increasing $a$ reduces this deviation.

The behaviors of the energy and angular momentum both exhibit a non-monotonic behavior with respect to the radial coordinate $r$: they first decrease and then increase as $r$ increases, with the energy asymptotically approaching 1 at spatial infinity. In particular, the extrema of the energy and angular momentum as functions of $r$ mark the location of the ISCO, which will be studied in detail in the next subsection as a preparation for our discussion of the structure of accretion disks in Sec. \ref{secDisk}. Furthermore, we find that in Figs. \subref{figpCircOrbitE2b}, \subref{figpCircOrbitL2c}, \subref{fignCircOrbitE2e}, and \subref{fignCircOrbitL2f}, the light solid curves always coincide with the dashed curves of the same color, indicating that the sign of the NUT charge has no effect on the values of the energy and angular momentum, whereas their values increase with the absolute value of the NUT charge.

\subsection{Innermost stable circular orbits}\label{subsecISCO}
As previously noted, the ISCO corresponds to the extremum of the energy or angular momentum of the circular orbits. In fact, this condition also applies to the innermost stable circular orbits (ISSOs). While this fact has been used in Refs. \cite{XiangChengMeng2025ImprintsOB,Wei,CChen} to determine the location of the ISCO or ISSO, to the best of our knowledge, a rigorous proof has not been provided. In the next section, we will present a proof of its equivalence to the conventional method of finding the ISCO or ISSO via the effective potential. Here, we adopt such property
\begin{equation}\label{eqrISCO}
	\left. \left( \frac{dE}{dr} \right) \right. _ {r=r_{\text{ISCO}},\theta_{0}=\theta_{0}^{\text{ISCO}},a,n}  =0,\qquad \left. \left( \frac{dL}{dr} \right) \right. _ {r=r_{\text{ISCO}},\theta_{0}=\theta_{0}^{\text{ISCO}},a,n}  =0
	\end{equation}
to obtain the ISCO. Obviously, for fixed $a$ and $n$, above equations involve the ISCO radius $r_{\text{ISCO}}$ and the angular position $\theta_0^{\text{ISCO}}$. While they are sufficient to determine both $r_{\text{ISCO}}$ and $\theta_0^{\text{ISCO}}$, the solution is sensitive to the choice of initial values. Therefore, in order to improve the numerical stability, we solve one of these equations by combining with Eq. \eqref{eqtheta0} .

\begin{figure}[!htbp]
    \centering{
			\subfigure[]{\includegraphics[width=7cm]{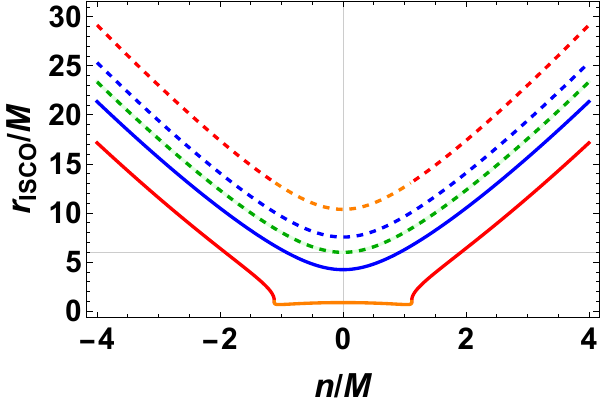}\label{figrISCOvsnNaked3a}}\qquad\qquad
			\subfigure[]{\includegraphics[width=7cm]{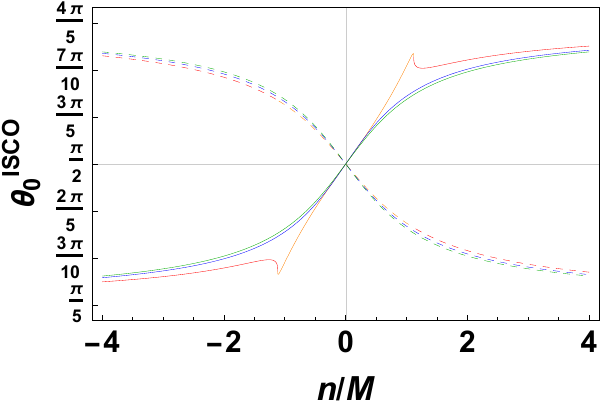}\label{figTheta0ISCOvsnNaked3b}}
			\caption{The variations of the ISCO radius $r_{\text{ISCO}}$ and angular position $\theta_0^{\text{ISCO}}$ with respect to the NUT charge $n$. The (light) green, blue, and red (orange) curves correspond to $a/M = 0$, $0.5$, and $1.5$, respectively. Solid and dashed curves represent prograde and retrograde orbits. For the curves with a = 1.5, the orange part corresponds to the naked singularity case, while the regions on both sides represent the RS-KTN black holes, marked in red. The junction of the two colors corresponds to the extremal RS-KTN black hole. (a) $r_{\text{ISCO}}$ vs $n$. (b) $\theta_0^{\text{ISCO}}$ vs $n$.
}\label{figISCO3}
    }
\end{figure}

In Fig.~\ref{figISCO3}, we show the influence of the NUT charge on the ISCO radius $r_{\text{ISCO}}$ and angular position $\theta_0^{\text{ISCO}}$ for different values of the spin $a$. The sign of the NUT charge does not influence the ISCO radius but does affect its angular position: similar to the general circular orbits, the sign of $nL$ determines whether the orbit lies in the northern or southern hemisphere. Furthermore, the greater the absolute value of the NUT charge $|n|$, the larger the ISCO radius and the greater its deviation from the equatorial plane. However, this behavior is violated in the cases of extremal black hole and naked singularity.

We also observe that retrograde orbits consistently have a larger radius than prograde ones. However, in the Taub-NUT spacetime ($a= 0$), these two orbits coincide, indicating that the spin $a$ causes the splitting between prograde and retrograde ISCOs. This suggests that although the presence of the NUT charge also yields a stationary spacetime like the Kerr case, it is fundamentally different from spin. Additionally, the spin $a$ has opposite effects on prograde and retrograde orbits: for the prograde ISCOs, a higher spin parameter $a$ leads to a smaller orbital radius and a greater deviation from the equatorial plane, whereas for the retrograde ISCOs, a higher $a$ results in a larger orbital radius and a smaller the angular deviation.

Notably, in the RS-KTN case, the prograde ISCO radius and angular position exhibit sharp corners near the extremal black hole, while no such feature appears for retrograde orbits. This may be attributed to the fact that the prograde ISCOs reside closer to the event horizon than the retrograde ones. As the black hole transitions from non-extremal to extremal and eventually to a naked singularity, the change in horizon structure affects the nearby geometry, thereby influencing the nearby the prograde ISCO.

\section{Precession of spherical orbits without reflection symmetry in \texorpdfstring{$\theta$}{theta} direction}\label{secPOSO}

We have seen that the presence of the NUT charge, which breaks the spacetime's $\mathbb{Z}_2$ symmetry, causes the circular orbits to deviate from the equatorial plane. Spherical orbits, which generalize circular orbits by allowing $\theta$ to vary, have a reflection symmetry about $\theta = \pi/2$ due to the $\mathbb{Z}_2$ symmetry of the spacetimes. Therefore, the introduction of the NUT charge breaks this reflection symmetry of the orbits in the $\theta$-direction as well. In this section, we investigate how the NUT charge influences the trajectories of the spherical orbits and examine its effect on the precession behavior of such orbits. Specifically, we first derive the equations of motion for the spherical orbits. Then we show that in the Taub-NUT spacetime, spherical orbits take the form of tilted circular orbits without precession. The precession behavior of the spherical orbits is investigated in the KTN spacetime.

Firstly, we derive the equations of motion for the spherical orbits whose circular orbits lie on the non-equatorial planes. Although Ref. \cite{XiangChengMeng2025ImprintsOB} derives the equations for the spherical orbits in general stationary axisymmetric spacetime, the analysis therein assumes unbroken $\mathbb{Z}_2$ symmetry. In contrast, our derivation is more general, as it includes spacetime without $\mathbb{Z}_2$ symmetry, such as the KTN spacetime. Following the strategy employed in the previous subsection for circular orbits, we begin by determining the conserved energy and angular momentum associated with the spherical orbits. To streamline the analysis and avoid redundant calculations, we proceed directly from Eq. \eqref{eqrconstraint} established in the preceding section.

Compared with circular orbits of the same orbital radius, spherical orbits are no longer confined to a fixed value of $\theta$; instead, they oscillate between $\theta_{\min}$ and $\theta_{\max}$. It is important to note that due to the breaking of symmetry in the $\theta$ direction, the midpoint between $\theta_{\min}$ and $\theta_{\max}$ typically does not coincide with $\theta_0$. We will elaborate on this feature in the next subsection. Now, we focus on the fact that $\theta_{\min}$ and $\theta_{\max}$ serve as turning points of the spherical orbit, where the constraint $\dot{\theta} = 0$ holds. These turning points allow us to obtain explicit expressions for the conserved energy and angular momentum.

    Without loss of generality, we choose $\theta_{\min} = \theta_0 - \zeta$ as a reference point to label different spherical orbits. Here, $\zeta$ is a characteristic angle that quantifies the deviation of the upper turning point from  the circular orbit with the same orbital radius as a spherical orbit. In the absence of the NUT charge, $\zeta$ reduces to the tilt angle introduced in Ref. \cite{XiangChengMeng2025ImprintsOB,Zahrani,Wei,CChen}. To distinguish it from the tilt angle, we temporarily refer to $\zeta$ as the deviation angle. As will be shown later, the difference between the two angles is negligible, and thus it is acceptable to treat the deviation angle as the tilt angle. At the turning point $\theta_{\min}$, we then obtain the condition
    \begin{equation}\label{eqturnpoint}
	\left. p_{\theta}\right| _{\theta =\theta_{0}-\zeta}  =0.
    \end{equation}
    Taking into account the constraints of the spherical orbit given in Eqs. \eqref{eqrconstraint} and \eqref{eqturnpoint}, Eqs. \eqref{eqprdot} and \eqref{eqtimelike} reduce to the following form
    \begin{align}
     \label{eqpthetadotSO}\left.\left(E^{2} \partial_{r}g^{tt} -2EL \partial_{r}g^{t\phi}+L^{2} \partial_{r}g^{\phi \phi}\right)\right|_{\theta =\theta_{0}-\zeta}&=0,\\
    \label{eqtimelikeSO}\left.\left(-g^{tt}E^2+2g^{t\phi}EL-g^{\phi \phi}L^2-1\right)\right|_{\theta =\theta_{0}-\zeta}&=0.
    \end{align}
    These two equations are structurally analogous to Eqs. \eqref{eqpthetadotCircular} and \eqref{eqtimelikeCircular}, differing only in the value of $\theta$. By solving them, we obtain the energy and angular momentum for the spherical orbits,
    \begin{align}
	\label{eqHESO}E=\left. \frac{1}{\sqrt{-g^{tt}-2g^{t\phi}\chi -g^{\phi \phi}\chi ^2}} \right|_{\theta =\theta_{0}-\zeta},\\
	\label{eqHLSO}L=\left. \frac{-\chi}{\sqrt{-g^{tt}-2g^{t\phi}\chi -g^{\phi \phi}\chi ^2}} \right|_{\theta =\theta_{0}-\zeta},
    \end{align}
    \begin{equation}
	\chi =\frac{p_{\phi}}{p_t}=\left. \frac{-\partial _rg^{t\phi}\pm \sqrt{\left( \partial _rg^{t\phi} \right) ^2-\partial _rg^{tt}\partial _rg^{\phi \phi}}}{\partial _rg^{\phi \phi}} \right|_{\theta =\theta_{0}-\zeta},
\end{equation}
    analogous to those of circular orbits.

    Overall, once the black hole's spin $a$, NUT charge $n$, orbital radius $r$, and the deviation angle $\zeta$ are specified, the corresponding circular orbit can then be determined by using the method outlined in the previous section, and then the energy and angular momentum will be determined. Subsequently, the trajectory can be obtained by solving the following equations
    \begin{align}
	\label{eqtdotmotionSO} \dot{t} &= -g^{tt}E + g^{t\phi}L,\\
	\label{eqphidotmotionSO} \dot{\phi} &=-g^{t\phi}E + g^{\phi \phi}L,\\
	\label{eqthetadotmotionSO} \dot{\theta}&= g^{\theta \theta}p_{\theta}, \\
	\label{eqpthetadotmotionSO} \dot{p}_{\theta}& = -\frac{1}{2}\left(E^{2} \partial_{\theta }g^{tt}-2EL\partial_{\theta}g^{t\phi} + L^{2} \partial_{\theta}g^{\phi \phi} +p_{\theta}^{2} \partial_{\theta}g^{\theta \theta} \right).
    \end{align}

\subsection{Energy, angular momentum and innermost stable spherical orbits}
To study the precession of the spherical orbits, it is necessary to solve their trajectories from the equations of motion. Before doing so, we must first obtain the two conserved quantities: energy and angular momentum. These are computed using Eqs.~\eqref{eqHESO} and \eqref{eqHLSO}, where the angular position $\theta_0$ of the corresponding circular orbit has been determined in Sec.~\ref{subseccircposi}. For the subsequent calculation to be relevant to astrophysics, we let the deviation angle $\zeta = 1.25^\circ$, the same as observed by EHT \cite{Cui}.

\begin{figure}[!htbp]
    \centering{
			\subfigure[]{\includegraphics[width=5.9cm]{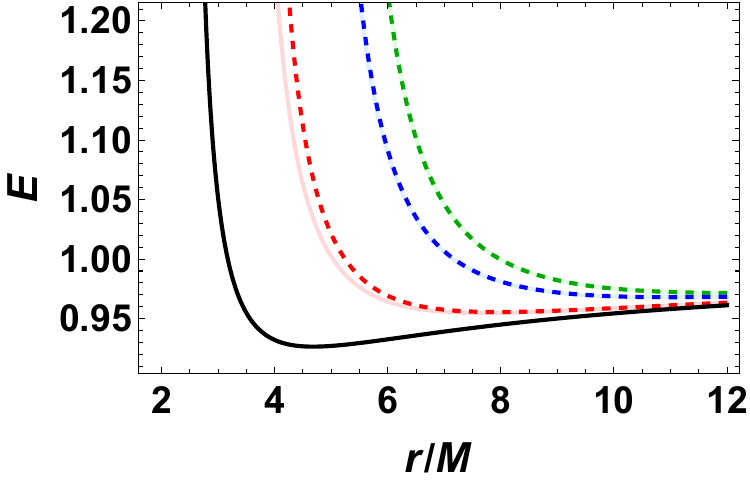}\label{figpSpheriOrbitE4a}}
			\subfigure[]{\includegraphics[width=5.7cm]{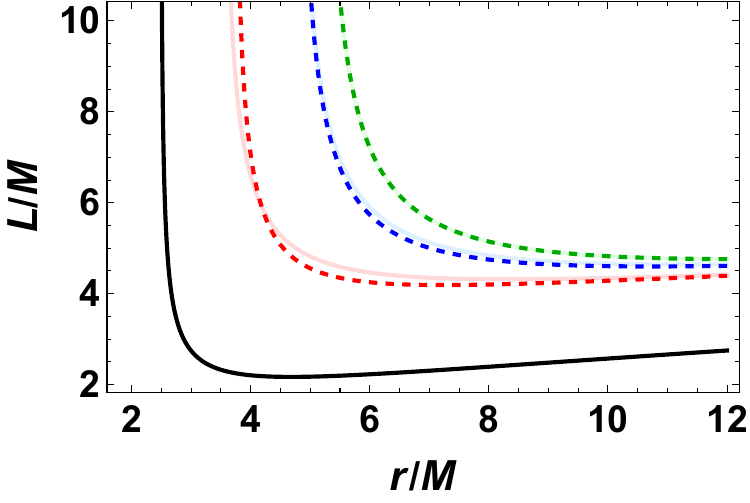}\label{figpSpheriOrbitL4b}}\\
            \subfigure[]{\includegraphics[width=5.9cm]{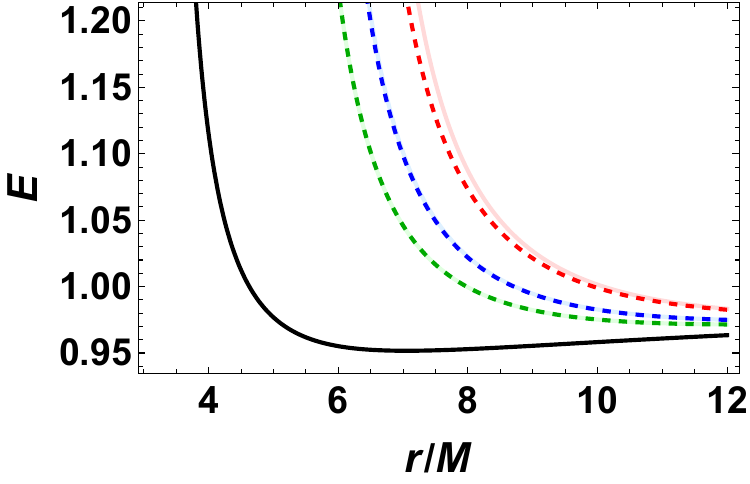}\label{fignSpheriOrbitE4c}}
            \subfigure[]{\includegraphics[width=5.9cm]{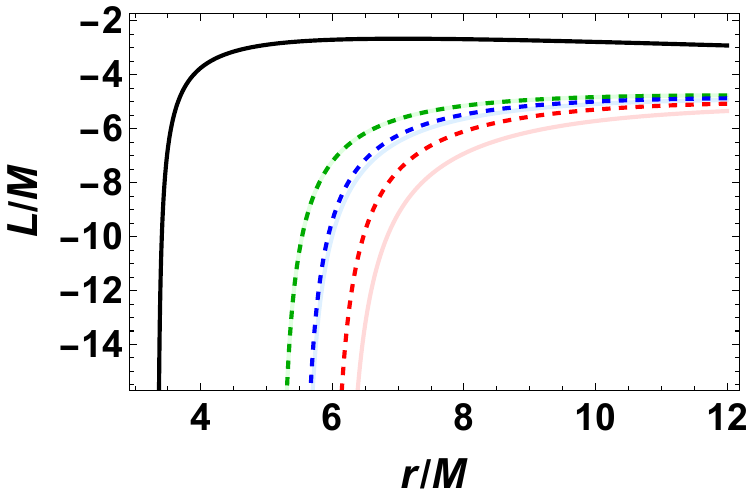}\label{fignSpheriOrbitL4d}}
			\caption{The energy $E$, and angular momentum $L$ of the spherical orbits vary with the radius $r$ in the Kerr, Taub-NUT, SS-KTN, and RS-KTN spacetimes. The deviation angle $\zeta$ is fixed at $1.25^\circ$. The upper and lower rows correspond to prograde and retrograde spherical orbits. The black solid curve represents the Kerr case with $a/M = 0.5$. The (light) green, (light) blue, and (light) red curves correspond to Taub-NUT, SS-KTN, and RS-KTN spacetimes with $a/M = 0$, $0.5$, and $1.5$, respectively. The dashed and solid curves denote the cases with $n/M = -2$ and $2$, respectively. (a) $E$ vs $r$ for prograde orbits. (b) $L$ vs $r$ for prograde orbits. (c) $E$ vs $r$ for retrograde orbits. (d) $L$ vs $r$ for retrograde orbits.}\label{figSOle4}
    }
\end{figure}

In Fig.~\ref{figSOle4}, we present the energy and angular momentum of the spherical orbits as functions of the orbital radius in the Taub-NUT, SS-KTN, and RS-KTN spacetimes. We find that the sign of the NUT charge has a slight effect on the energy and angular momentum, which diminishes as the radius $r$ increases but becomes more pronounced with increasing spin $a$. This asymmetry arises because the spherical orbits with positive and negative NUT charges are not located symmetrically on the sphere: while circular orbits are symmetric about the equatorial plane, spherical orbits with the same tilt angle are generally not. As a result, their energy and angular momentum differ. Moreover, the larger the absolute value of the NUT charge, the greater the energy and the absolute value of the angular momentum of the spherical orbits.

Similar to the circular orbits, the value of energy and the absolute value of the angular momentum for the spherical orbits first decrease and then increase as the radius increases, with the energy approaching 1 at spatial infinity. Since the spherical orbits generalize circular orbits, the ISSOs serve as the extension of the ISCOs. Consequently, the extrema of the energy and angular momentum as functions of radius correspond to the ISSO. Hence, the ISSO satisfies the condition given by
\begin{equation}\label{eqrISSO}
    \left. \left( \frac{dE}{dr} \right) \right. _ {r=r_{\text{ISSO}}, \theta_{0}-\zeta,a,n}  =0,\qquad \left. \left( \frac{dL}{dr} \right) \right. _ {r=r_{\text{ISSO}}, \theta_{0}-\zeta,a,n}  =0.
\end{equation}
Note that $\theta_0$ here is generally not equal to $\theta_0^{\text{ISCO}}$.

Before calculating the radius of the ISSO, we first prove this equation as well as Eq.~\eqref{eqrISCO} in Sec.~\ref{subsecISCO}. Specifically, we show that they are consistent with the standard method of determining the ISCOs and ISSOs by using the effective potential approach. Typically, the radial effective potential is used to characterize the radial motion of a particle, and is obtained by separating variables in the Hamilton-Jacobi method \cite{Carter,Wilkins,Dymnikova}. To ensure the generality of our proof, we express the radial effective potential $R$ as a function of the orbital radius $r$, the energy $E$, the angular momentum $L$, the angular position $\theta_0$ of circular orbit, the deviation angle $\zeta$, and the black hole parameters collectively denoted by $\alpha$, i.e.,
\[R(r, E, L, \theta_0, \zeta, \alpha).
\]
Here, $\alpha$ includes a set of black hole parameters such as the spin $a$ and the NUT charge $n$. The angular position $\theta_0$ takes the value $\pi/2$ in spacetimes with $\mathbb{Z}_2$ symmetry, and in spacetimes with a nonvanishing NUT charge, its value can be determined from the results of the previous section. The deviation angle $\zeta$ labels the spherical orbits: it is zero for circular orbits and set to a fixed value for spherical ones. Additionally, the black hole parameters $\alpha$ must be specified before obtaining the innermost stable orbit. Therefore, when evaluating the ISCO or ISSO, only the variables $r$, $E$, and $L$ vary, while the other parameters are kept fixed.

For the circular or spherical orbits, the effective potential must satisfy the following conditions
\begin{align}
    \label{eqEffR}R\left( r,E,L ,\theta _0,\zeta ,\alpha\right) =0, \\
    \label{eqpdEffR}\partial _rR\left( r,E,L ,\theta _0,\zeta ,\alpha\right) =0.
\end{align}
Furthermore, the innermost stable orbit corresponds to an inflection point of the effective potential \cite{Ruffini,Bardeen,Wilkins,Dymnikova,Zahrani}, which implies
\begin{equation}\label{eqpdpdEffR}
\partial _{rr}R\left( r,E,L ,\theta _0,\zeta ,\alpha\right) =0.
\end{equation}
By combining above three equations, one can directly solve for the radius, energy, and angular momentum of the innermost stable orbit.

Alternatively, we first solve Eqs.~\eqref{eqEffR} and \eqref{eqpdEffR} together to obtain the energy and angular momentum as functions of $r$
\begin{equation}
    \tilde{E}\left(r, \theta_0, \zeta, \alpha\right), \quad \tilde{L}\left(r, \theta_0, \zeta, \alpha\right).
\end{equation}
This step is equivalent to obtain the analytical expressions for energy and angular momentum in the Hamiltonian formalism, as given in Eqs.~\eqref{eqHECirc}, \eqref{eqHLCirc}, \eqref{eqHESO}, and \eqref{eqHLSO}.
We then substitute $E = \tilde{E}$ and $L = \tilde{L}$ back into the left-hand sides of Eqs.~\eqref{eqEffR} and \eqref{eqpdEffR}, which yields two functions of $r$ that identically vanish, i.e.,
\begin{align}
    R\left( r,\tilde{E},\tilde{L} ,\theta _0,\zeta ,\alpha\right) \equiv 0,\\
    \partial _rR\left( r,\tilde{E},\tilde{L} ,\theta _0,\zeta ,\alpha\right) \equiv 0.
\end{align}
Consequently, the total differential of the left-hand side with respect to $r$ also vanishes
\begin{align}
    \frac{d}{dr}R\left( r,\tilde{E},\tilde{L},\alpha ,\theta _0,\zeta \right) =R^{\left( 1,0,0 \right)}+R^{\left( 0,1,0 \right)}\frac{d\tilde{E}}{dr}+R^{\left( 0,0,1 \right)}\frac{d\tilde{L}}{dr}\equiv 0,\\
    \frac{d}{dr}\partial _rR\left( r,\tilde{E},\tilde{L},\alpha ,\theta _0,\zeta \right) =R^{\left( 2,0,0 \right)}+R^{\left( 1,1,0 \right)}\frac{d\tilde{E}}{dr}+R^{\left( 1,0,1 \right)}\frac{d\tilde{L}}{dr}=0,
\end{align}
where the superscripts of $R$ represent partial derivatives with respect to $r$, $\tilde{E}$, and $\tilde{L}$, respectively.

This clearly forms a system of two inhomogeneous linear equations for $\frac{\mathrm{d}\tilde{E}}{\mathrm{d}r}$ and $\frac{\mathrm{d}\tilde{L}}{\mathrm{d}r}$. Assuming the determinant of the coefficient matrix is nonzero, solving this system yields explicit expressions for these derivatives
\begin{align}
    \frac{d\tilde{E}}{dr}&=\frac{R^{\left( 1,0,0 \right)}R^{\left( 1,0,1 \right)}-R^{\left( 0,0,1 \right)}R^{\left( 2,0,0 \right)}}{R^{\left( 0,0,1 \right)}R^{\left( 1,1,0 \right)}-R^{\left( 0,1,0 \right)}R^{\left( 1,0,1 \right)}},\\
    \frac{d\tilde{L}}{dr}&=\frac{R^{\left( 1,0,0 \right)}R^{\left( 1,1,0 \right)}-R^{\left( 0,1,0 \right)}R^{\left( 2,0,0 \right)}}{R^{\left( 0,1,0 \right)}R^{\left( 1,0,1 \right)}-R^{\left( 0,0,1 \right)}R^{\left( 1,1,0 \right)}},
\end{align}
where the arguments of $R$ are omitted here. From Eqs.~\eqref{eqpdEffR} and \eqref{eqpdpdEffR}, we have
\begin{equation}
R^{(1,0,0)} = R^{(2,0,0)} = 0,
\end{equation}
which leads to
\begin{equation}
\frac{d\tilde{E}}{dr} = \frac{d\tilde{L}}{dr}=0.
\end{equation}
As a result, starting from the effective potential, we have demonstrated that the energy and angular momentum attain extrema at the innermost stable orbit. This lays a solid foundation for determining the radius and angular position of the ISCO and ISSO.

\begin{figure}[!htbp]
    \centering{
			\subfigure[]{\includegraphics[width=7cm]{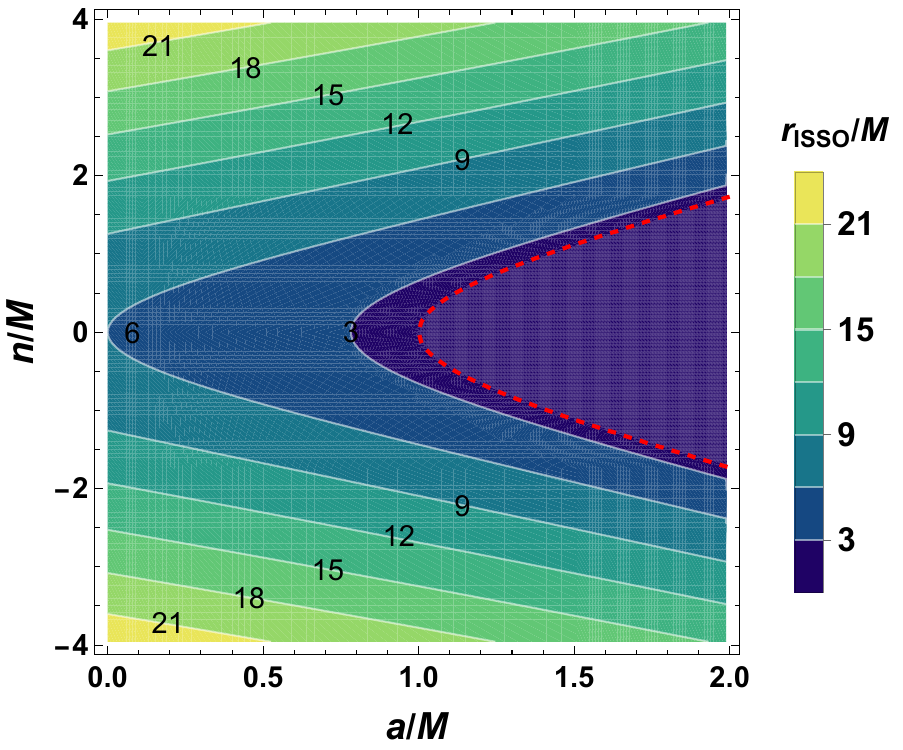}\label{figprISSO5a}}
			\subfigure[]{\includegraphics[width=7cm]{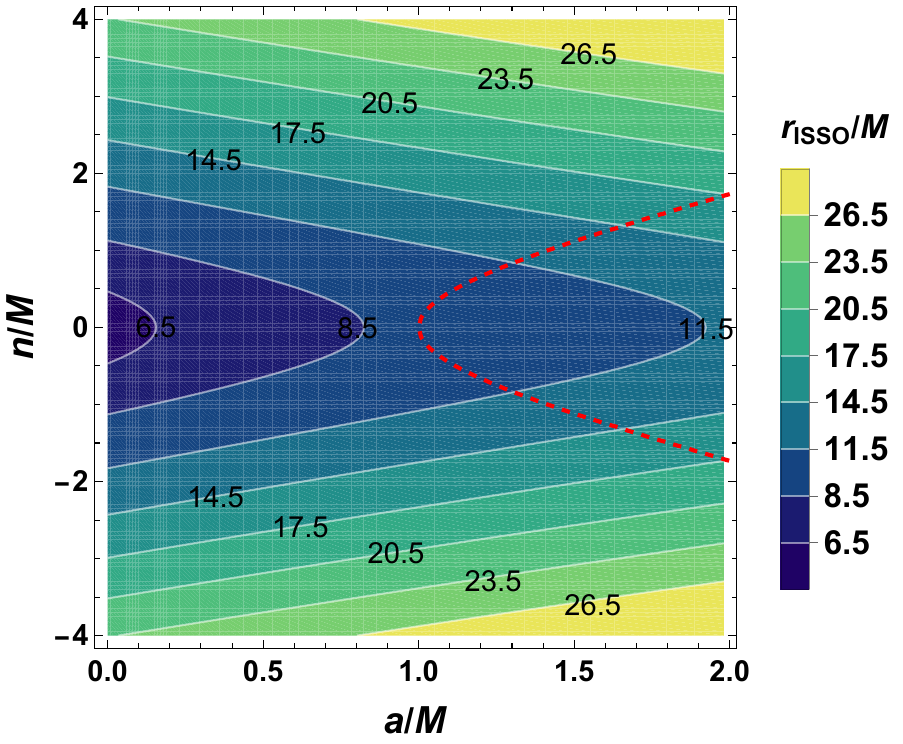}\label{fignrISSO5b}}
			\caption{The orbital radius $r_{\text{ISSO}}$ of the ISSO as a function of the spin $a$ and NUT charge $n$, with a deviation angle $\zeta = 1.25^\circ$. The red dashed curve represents the extremal RS-KTN black hole, and the region enclosed by it corresponds to the naked singularity regime. (a) Prograde orbits. (b) Retrograde orbits.
}\label{figISSO5}
    }
\end{figure}

At the end of this subsection, we solve one of Eq.~\eqref{eqrISSO} together with Eq.~\eqref{eqtheta0} to obtain the ISSO radius $r_{\text{ISSO}}$ and the  angular position $\theta_0$ of the corresponding circular orbit. In Fig.~\ref{figISSO5}, we illustrate how $r_{\text{ISSO}}$ varies with the spin $a$ and the NUT charge $n$. Although the sign of the NUT charge affects the energy and angular momentum of the spherical orbits, it has no influence on the ISSO radius; thus, both plots are symmetric with respect to $n = 0$. This symmetry is also evident in Fig.~\ref{figSOle4}, where the dashed and solid curves coincide to the right of the extremum. Moreover, we observe that a larger absolute value of the NUT charge leads to a larger $r_{\text{ISSO}}$. In addition, the prograde and retrograde cases exhibit different behaviors in the naked singularity region. For prograde orbits, $r_{\text{ISSO}}$ in this region is always smaller than $2M$, and the entire region is enclosed by a contour near $2M$. In contrast, for retrograde orbits, $r_{\text{ISSO}}$ in this region is always greater than $8M$, and this region is connected to the black hole region in parameter space. These behaviors resemble those observed in Fig.~\ref{figISCO3}, possibly because the innermost stable prograde orbits are located closer to the event horizon than the retrograde ones.

\subsection{Tilted circular orbits in the Taub-NUT spacetime}\label{subsecTNSO}
Building on the previously derived equations of motion for the circular and spherical orbits, this subsection delves into the study of the spherical orbits in the Taub-NUT spacetime. To illustrate their properties concretely, we fix the orbital radius at $r = 10M$ and set the deviation angle to $\zeta = \pi/4$, then compute the corresponding trajectories numerically.

\begin{figure}[!htbp]
    \centering{
			\subfigure{\includegraphics[width=7cm]{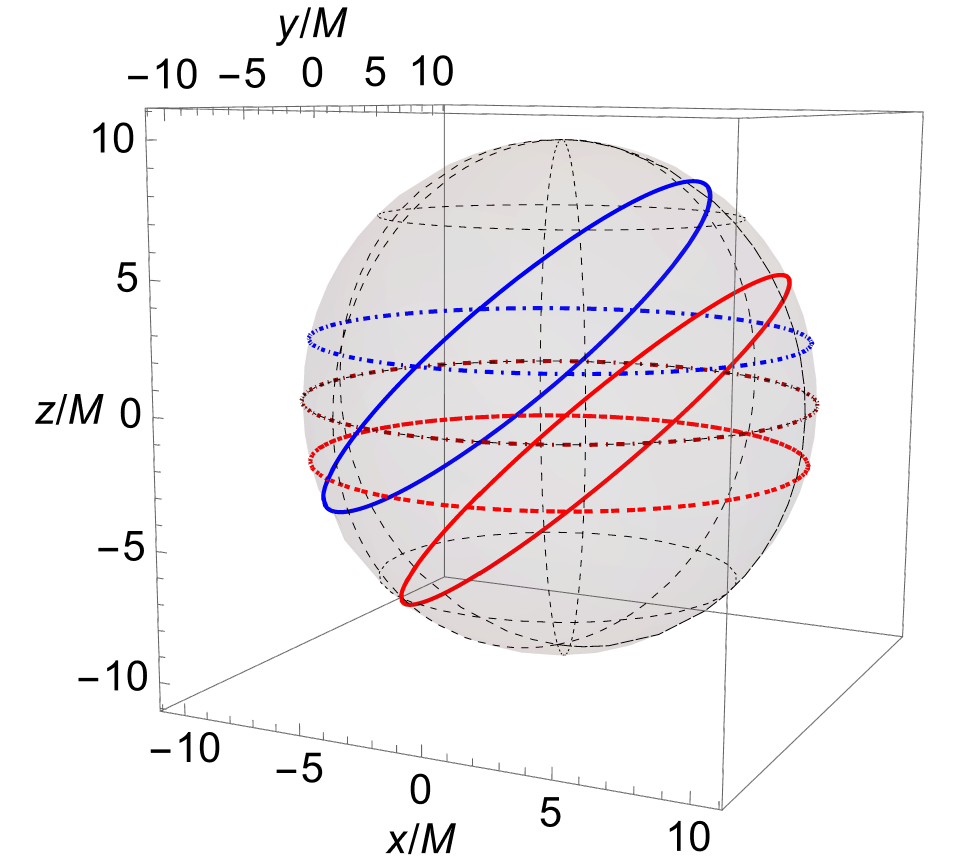}}
			\caption{The trajectories of prograde tilted circular orbits with $r/M = 10$, $\zeta = \pi/4$ in the Taub-NUT spacetime. The gray sphere represents the spherical surface on which the tilted circular orbits lie. The brown dashed circle marks the equator. The red and blue solid tilted rings represent the trajectories of tilted circular orbits with NUT charges $n = 0.5M$ and $n = -0.5M$, respectively. The red and blue dashed rings correspond to their non-equatorial circular orbits.
}\label{figNUTSphericalOrbit6}
    }
\end{figure}

We begin by determining the angular position $\theta_0$ of the circular orbit by using Eq.~\eqref{eqtheta0}. This value is then substituted into Eqs.~\eqref{eqHECirc} and \eqref{eqHLCirc} to evaluate the corresponding energy and angular momentum. The resulting circular trajectories are obtained through numerical integration of Eqs.~\eqref{eqCirctdot} and \eqref{eqCircphidot}. As depicted in Fig.~\ref{figNUTSphericalOrbit6}, the red and blue dashed rings, both lying parallel to the equator, represent prograde circular orbits with NUT charges $n = 0.5M$ and $-0.5M$, respectively. Notably, when $n > 0$, the orbit resides below the equatorial plane, while for $n < 0$, it lies above, in agreement with the result in Sec.~\ref{subseccircposi}.

To study the associated spherical orbits, we substitute $r = 10M$ and the previously obtained $\theta_0$ into Eqs.~\eqref{eqHESO} and \eqref{eqHLSO} to obtain the energy and angular momentum, and then solve Eqs.~\eqref{eqtdotmotionSO}-\eqref{eqpthetadotmotionSO} to compute the trajectories. As shown in Fig.~\ref{figNUTSphericalOrbit6}, the resulting orbits appear as tilted circular rings relative to the circular orbits. For clarity, we refer to these as tilted circular orbits.

Although the NUT charge makes the spacetime stationary, its symmetry is richer than that of the Kerr-like spacetime. From the viewpoint of these tilted orbits, the NUT spacetime behaves more like a stationary spherically symmetric spacetime. In fact, Misner showed that the Taub-NUT metric is invariant under a globally well-defined $\mathrm{SO}(3)$ action \cite{Misner1963}.

In addition, these tilted orbits are closed, indicating the absence of precession. This demonstrates that the NUT charge $n$ does not induce the LT precession in the way the spin $a$ does. A rigorous proof of this result can be found in Ref.~\cite{VKagramanova}. Overall, the so-called ``spherical orbits'' in the Taub-NUT spacetime are in fact tilted circular orbits without precession.

\subsection{Precession of spherical orbits in Kerr-Taub-NUT spacetime}

Since the spherical orbits in spacetimes with only a nonzero NUT charge appear as tilted circular orbits without precession, introducing a nonvanishing spin $a$ is necessary to investigate the influence of the NUT charge on the precession. In this subsection, we study the characteristics and precession behavior of the spherical orbits in the KTN spacetime. Due to the presence of the NUT charge, the $\mathbb{Z}_2$ symmetry of the spacetime is broken, which in turn leads to a loss of reflection symmetry about the circular orbit in the $\theta$-motion. We analyze the properties of these orbits by examining their equations of motion and solving them numerically.

\begin{figure}[!htbp]
    \centering{
			\subfigure[]{\includegraphics[width=5.5cm]{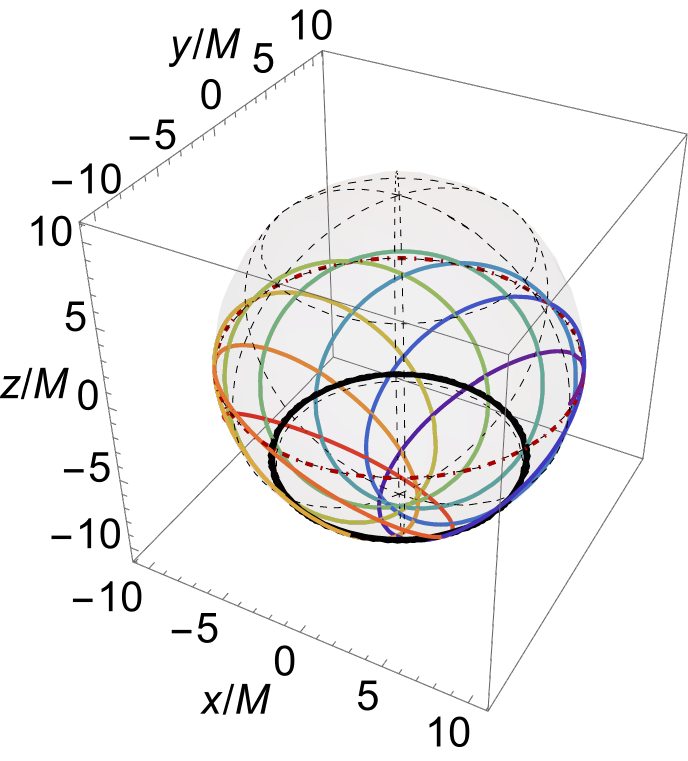}\label{figTrajectory3D7a}}\\
			\subfigure[]{\includegraphics[width=5.1cm]{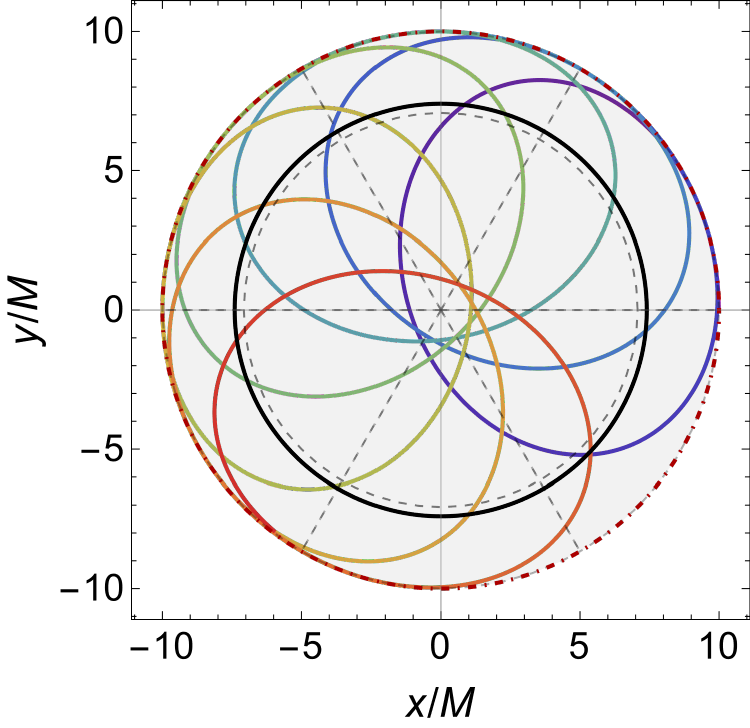}\label{figTrajectoryXY7b}}
            \subfigure[]{\includegraphics[width=5.1cm]{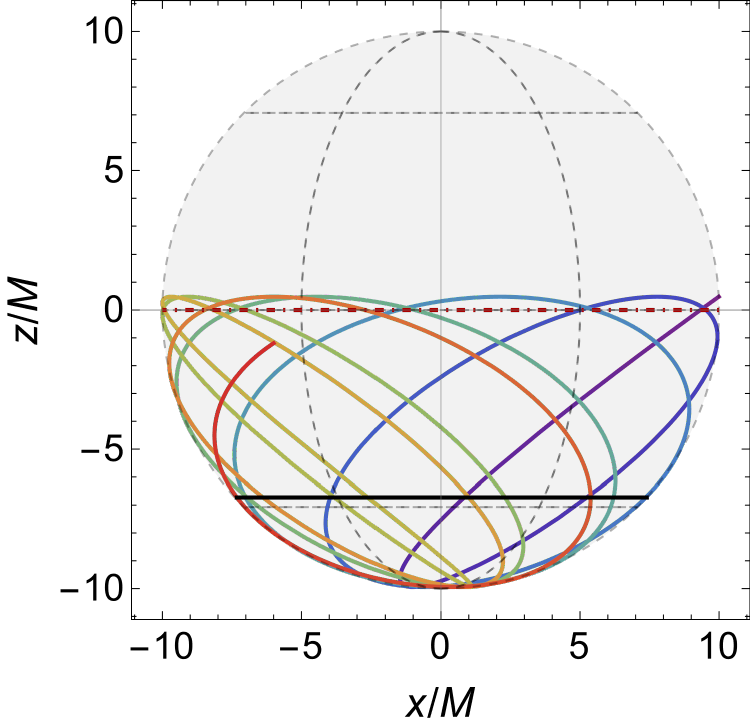}\label{figTrajectoryXZ7c}}
			\caption{The trajectory of a prograde spherical orbit with radius $r/M = 10$ and deviation angle $\zeta = \pi/4$ in the RS-KTN spacetime with $a/M = 1.5$ and $n/M = 4$. The gray sphere represents the spherical surface on which the orbit lies. The brown dashed ring denotes the equator. The black solid ring corresponds to the associated circular orbit. The rainbow-colored curve illustrates the trajectory of the spherical orbit, with the color gradually transitioning from dark blue to dark red as the proper time $\tau$ increases. (a) 3-dimensional view of the spherical orbit. (b) Projection of the orbit onto the $xy$-plane. (c) Projection of the orbit onto the $xz$-plane.
}\label{figTrajectory7}
    }
\end{figure}

Following the same procedure as in the previous subsection for the Taub-NUT case, we compute the trajectory of the spherical orbits in the KTN spacetime. In Fig.~\ref{figTrajectory7}, we present the trajectory of a prograde spherical orbit with radius $r = 10M$ in the RS-KTN spacetime. To emphasize the deviation of the circular orbit from the equatorial plane, the relative inclination of the spherical orbit, and to amplify the precession effect, we set $n = 4M$, $\zeta = \pi/4$, and $a = 1.5M$. In Fig.~\ref{figTrajectory3D7a}, the 3-dimensional trajectory is shown: the circular orbit (black solid ring) lies below the equatorial plane, while the spherical orbit (rainbow-colored curve) is tilted with respect to it. As will be shown later, this tilt angle can be well approximated by the deviation angle~$\zeta$. Therefore, the presence of the NUT charge breaks the reflection symmetry of the spherical orbits, so they no longer exhibit symmetry about equatorial plane as in the Kerr case. This asymmetry is also clearly visible in the $xz$-plane projection in Fig.~\ref{figTrajectoryXZ7c}. Moreover, the spherical orbit is not closed, indicating the presence of a precession effect. From Fig.~\ref{figTrajectoryXY7b}, which shows the projection onto the $xy$-plane, it is evident that the orbital plane gradually rotates around the $z$-axis as proper time evolves.

\begin{figure}[!htbp]
    \centering{
			\subfigure[]{\includegraphics[width=8.7cm]{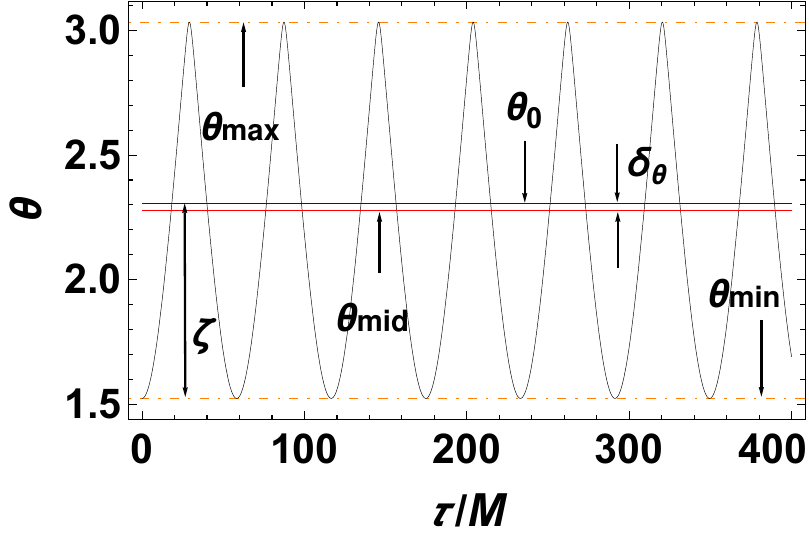}\label{figThetaMotion8a}}
            \subfigure[]{\includegraphics[width=9cm]{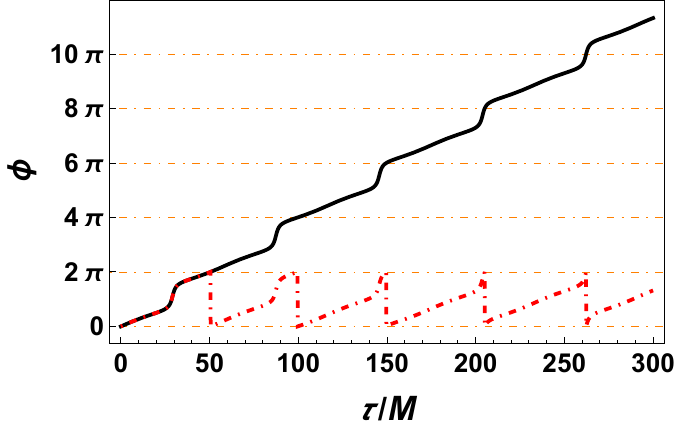}\label{figPhiMotion8b}}
			\caption{The evolution of the coordinates $\theta$ and $\phi$ with respect to the proper time $\tau$ for the spherical orbit with $r/M = 10$ and $\zeta = \pi/4$ shown in Fig.~\ref{figTrajectory7},  in the RS-KTN spacetime with $a/M = 1.5$ and $n/M = 4$.
(a) The black solid curve shows the evolution of $\theta$ as a function of proper time $\tau$. The upper and lower orange dot-dashed lines represent the maximum and minimum values of $\theta$, denoted as $\theta_{\text{max}}$ and $\theta_{\text{min}}$. The red and blue lines indicate the midpoint $\theta_{\text{mid}} = \frac{\theta_{\text{min}} + \theta_{\text{max}}}{2}$ and the angular position $\theta_0$ of the corresponding circular orbit, respectively. The difference between $\theta_{\text{mid}}$ and $\theta_0$ is denoted by $\delta_\theta$. The difference between the upper turning point $\theta_{\text{min}}$ and the angular position $\theta_0$ of circular orbit is the deviation angle $\zeta$, whereas the usual tilt angle is given by $\frac{\theta_{\text{max}} - \theta_{\text{min}}}{2}$.
(b) The black solid curve shows the cumulative evolution of the azimuthal angle $\phi$ as a function of $\tau$, obtained by solving the equations of motion. Since the physical motion of $\phi$ is confined to the interval $[0, 2\pi)$, the red dot-dashed curve shows the true angular trajectory with the accumulated $2\pi$ rotations subtracted.
}\label{figMotion8}
    }
\end{figure}

To clearly illustrate the breaking of symmetry in the spherical orbits, in Fig.~8 we plot the evolution of $\theta$ and $\phi$ with respect to the proper time $\tau$ for the spherical orbit shown in Fig.~\ref{figTrajectory7}. In Fig. \ref{figThetaMotion8a}, we display the evolution of $\theta$, where a small angular difference $\delta_\theta$ can be observed between the angular position $\theta_0$ of the corresponding circular orbit and the midpoint $\theta^{\text{mid}}$. Although the motion in the $\theta$-direction is periodic, we observe that the peaks are slightly sharper than the troughs within each cycle, indicating a violation of reflection symmetry about the circular orbit. Specifically, under the transformation $\tau \rightarrow 2\tau_{0}-\tau$, $\theta \rightarrow 2\theta_0 - \theta$, where $\tau_{0}$ represents the moment when the particle is in the circular orbit, the evolution curve of $\theta$ does not remain invariant.

We can further confirm this asymmetry about equatorial plane by analyzing the equations of motion. If the motion in the $\theta$-direction possessed reflection symmetry about the circular orbit, then the evolution curves of $\theta$ and $p_\theta$ would resemble sinusoidal functions, and should satisfy the following condition
\begin{equation}\label{eqthetaptehtadot}
    \dot{\theta}|_{\theta _0-\beta}=\dot{\theta}|_{\theta _0+\beta},\qquad p_{\theta}|_{\theta _0-\beta}=p_{\theta}|_{\theta _0+\beta}.
\end{equation}
Here, $\beta$ denotes a small angle relative to $\theta_0$ which is the center of symmetry. When $\beta = \zeta$, the two sides of the equation correspond to the values of the function at the trough and the peak, respectively. From Eq. \eqref{eqthetaptehtadot},  we have the following relation
\begin{equation}
    \frac{\left( \frac{\dot{\theta}}{p_{\theta}} \right) |_{\theta _0-\beta}}{\left( \frac{\dot{\theta}}{p_{\theta}} \right) |_{\theta _0+\beta}}=1.
\end{equation}
Using Eq.~\eqref{eqthetadotmotionSO}, we obtain
\begin{equation}
    \frac{g^{\theta \theta}|_{\theta _0-\beta}}{g^{\theta \theta}|_{\theta _0+\beta}}=1.
\end{equation}
For the KTN black holes, we have
\begin{equation}
    \frac{g^{\theta \theta}|_{\theta _0-\beta}}{g^{\theta \theta}|_{\theta _0+\beta}}=\frac{r^2+\left( n+a\cos \left( \theta _0-\beta \right) \right) ^2}{r^2+\left( n+a\cos \left( \theta _0+\beta \right) \right) ^2}\neq1,
\end{equation}
unless $n=0$ and $\theta_0 = \pi/2$. However, as we have shown in Sec.~\ref{subseccircposi}, when the NUT charge $n$ is nonzero, the circular orbit does not lie in the equatorial plane, i.e., $\theta_0 \neq \pi/2$. Therefore, for $n \neq 0$, the $\theta$-motion cannot possess reflection symmetry about $\theta_0$.

\begin{figure}[!htbp]
    \centering{
			\subfigure{\includegraphics[width=9cm]{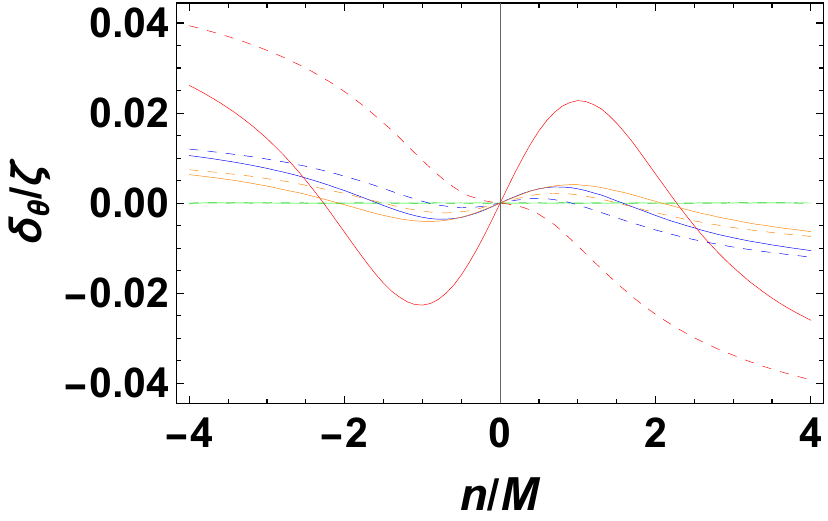}}
			\caption{Variation of the ratio $\delta_{\theta}/\zeta$ with the NUT charge $n$. The deviation angle $\zeta$ is set at $1.25^\circ$. The (dark) green, blue, red, and orange curves represent cases with $a/M = 0$, $0.5$, $1.5$, and $0.5$, respectively. The orbital radius is $r/M = 15$ for all cases except the orange curve, which corresponds to $r/M = 20$. Solid curves denote the prograde orbits, while dashed curves denote the retrograde ones.
}\label{figAngularShift9}
    }
\end{figure}

Given that the angular position $\theta_0$ of circular orbit is not the symmetry center of the $\theta$-motion, it is important to assess its deviation from the midpoint $\theta_{\text{mid}}$, as this determines whether the tilt angle of the spherical orbit can be approximated by the deviation angle $\zeta$. The tilt angle is typically defined as $\frac{\theta_{\text{max}} - \theta_{\text{min}}}{2}$ \cite{Zahrani,Wei,CChen,XiangChengMeng2025ImprintsOB}. To justify the approximation, we examine the difference between $\theta_0$ and $\theta_{\text{mid}}$, denoted by $\delta_\theta$ in Fig.~\ref{figThetaMotion8a}, and evaluate whether it is negligible compared to $\zeta$. In Fig.~\ref{figAngularShift9}, we show the ratio $\delta_\theta / \zeta$ as a function of the NUT charge $n$, under various values of the orbital radius $r$, deviation angle $\zeta$, and spin parameter $a$. The results show that $\delta_{\theta}$ is always less than 4\% of $\zeta$ throughout the parameter range we explored. Moreover, the ratio $|\delta_{\theta}/\zeta|$ generally increases with $|n|$ but decreases with $r$. As shown in Fig.~\ref{figISSO5}, the radius $r_{\text{ISSO}}$ grows with $|n|$, and in the next section we will focus on the spherical orbits with $r > r_{\text{ISSO}}$. This implies that in the parameter region of interest, the effects of increasing $|n|$ and $r$ tend to counteract each other, preventing $|\delta_{\theta}/\zeta|$ from becoming too large. Therefore, the deviation angle $\zeta$ serves as a good approximation for the tilt angle of the spherical orbits relative to the associated circular orbits. To avoid the complexity of terminology and to maintain consistency with previous literature, we will refer to $\zeta$ as the tilt angle in the following discussion, while keeping in mind the subtle distinction between the deviation angle and the tilt angle.

We now turn to the calculation of the precession angular velocity of the spherical orbits in the KTN spacetime and investigate how it is affected by the NUT charge. To highlight this precession effect, the red dotted curve in Fig.~\ref{figPhiMotion8b} shows the actual evolution of the $\phi$ coordinate, obtained by numerically solving the equations of motion and removing the $2\pi$-periodic accumulation. By comparing Fig.~\ref{figThetaMotion8a} and \subref{figPhiMotion8b}, we observe that the periods of the $\theta$ and $\phi$ motions are not equal. As a result, the orbit is not closed and exhibits precession. Specifically, we use the following formula to compute the precession angular velocity as seen by a distant observer
\begin{equation}\label{eqomegat}
    \omega_{\text{t}} = \frac{\Delta\phi \mp 2\pi}{T_{\theta}},
\end{equation}
where $T_{\theta}$ is the period of $\theta$-motion measured in coordinate time, obtained by solving Eq.~\eqref{eqtdotmotionSO}, and $\Delta\phi$ is the increment of the $\phi$ during one full $\theta$-oscillation. The sign ``$\mp$" corresponds to prograde and retrograde orbits, respectively. For details of the computational procedure, please refer to Ref. \cite{XiangChengMeng2025ImprintsOB}.

\begin{figure}[!htbp]
    \centering{
			\subfigure{\includegraphics[width=9cm]{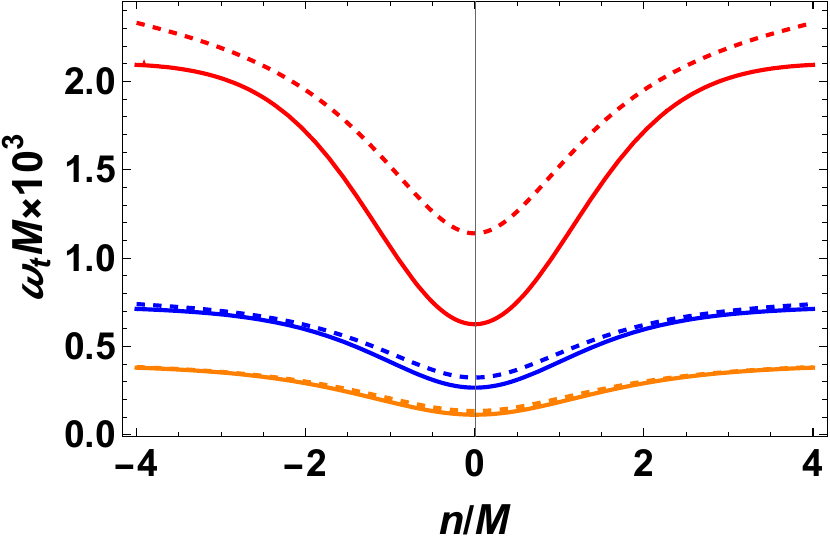}}
			\caption{
        The precession angular velocity $\omega_{t}$ of the spherical orbits in the KTN spacetime as a function of the NUT charge $n$. The tilt angle $\zeta$ is set at $1.25^\circ$. Solid and dashed curves correspond to prograde and retrograde orbits, respectively. The blue, red, and orange curves represent $a/M = 0.5$, $1.5$, and $0.5$, respectively, with the corresponding orbital radii $r/M = 15$, $15$, and $20$.
}\label{figomegat10}
    }
\end{figure}

The precession angular velocity $\omega_{t}$ of the spherical orbits is plotted in Fig.~\ref{figomegat10} as a function of the NUT charge $n$, with the tilt angle $\zeta$ fixed at $1.25^\circ$. We observe that $\omega_{t}$ is symmetric with respect to $n=0$, indicating that the sign of the NUT charge does not affect the precession. As $|n|$ increases, the precession angular velocity $\omega_{t}$ increases accordingly. This trend holds for both prograde and retrograde orbits and for different values of the spin parameter $a$. It is important to note that the sign of the NUT charge can slightly affect the energy and angular momentum of the spherical orbits with radii smaller than $r_{\text{ISSO}}$. As a result, the precession angular velocity $\omega_{t}$ of such orbits may also exhibit a weak dependence on the sign of $n$. However, since these orbits are not of primary relevance for the astrophysical applications discussed in the next section, we will not explore the spherical orbits with $r < r_{\text{ISSO}}$ in detail.

\section{A tilted  accretion  disk around a Kerr-Taub-NUT black hole}\label{secDisk}
So far, we have explored the rich features of the spherical orbits in the KTN spacetime induced by the presence of the NUT charge. First, the emergence of non-equatorial circular orbits broadens the study beyond equatorial motion, making the determination of the angular position $\theta_0$ of these orbits a prerequisite for further investigations of spherical trajectories. Second, the existence of non-precessing tilted circular orbits in the Taub-NUT spacetime reveals the distinct roles of the NUT charge $n$ and the spin parameter $a$. Moreover, the NUT charge breaks the $\mathbb{Z}_2$ symmetry of the spacetime, which manifests in the spherical orbits as a violation of the symmetry under simultaneous space-time reflection about the circular orbit in the $\theta$ direction. Finally, we observed that increasing the absolute value $|n|$ leads to an enhancement of the precession angular velocity $\omega_t$. Given the rich and significant influence of the NUT charge on the spherical orbits, it is expected to have a non-negligible impact on models based on such orbits, particularly those related to astrophysical observations. In fact, astronomical observations suggest that the NUT charge, as a gravitomagnetic monopole, may indeed exist~\cite{Chakraborty2018,Narzilloev2023,Sen2024,Ghasemi2021}.

In recent years, the observations of the black hole shadow image of M87* and SgrA* attracted widespread attention~\cite{Akiyama1,Akiyama2}. As a manifestation of strong-field gravity, black hole shadows have emerged as an effective probe for testing gravity theories and constraining on black hole parameters. Following the shadow observation, recent measurements of jet precession in M87* have sustained interest in black holes and their accretion disks. Based on 22 years of radio observations, Cui et al. reported an 11-year precession period of the black hole jet, which is possibly driven by LT precession due to a spinning black hole with a misaligned accretion disk~\cite{Cui}. Subsequently, Wei et al. were the first to use this observation to constrain the spin of a Kerr black hole ~\cite{Wei}. Cui and Lin examined the impact of M87's jet precession on the black hole-accretion disk system, including its influence on the disk size and the jet's non-collinear morphology ~\cite{Cui2}. Iorio reproduced the temporal pattern and amplitude of the jet orientation using perturbative calculations at the first post-Newtonian order ~\cite{Iorio}. In addition, the precession of the spherical orbits has also been investigated in other black hole spacetimes to constrain black hole parameters ~\cite{CChen,WangHuiMin,WangHuiMin2507}.

Very recently, we investigated the precession of the spherical orbits using the Hamiltonian formalism and further explored the capability of jet precession to constrain multiple parameters by taking the Kerr-Newman black hole as an example~\cite{XiangChengMeng2025ImprintsOB}. The results indicate that, although it does not yield precise values of the parameters, it can rule out certain regions of the parameter space and offer insights for accurate general relativistic magnetohydrodynamics (GRMHD) simulations. Furthermore, when combined with other observational constraints, the parameter space can be further tightened. Motivated by these developments, in this section we employ the spherical orbits in the KTN spacetime to model the motion of particles on the tilted accretion disk of M87*, thereby constraining the allowed parameter space of the spin $a$ and NUT charge $n$, and investigating their impact on the structure of the disk.

In previous studies \cite{BardeenPetterson,Petterson,Ostriker,Fragile,LodatoPrice}, the tilt angle of a misaligned accretion disk typically decreases with the orbital radius, eventually aligning with the equatorial plane at a characteristic radius known as the warp radius. From the warp radius inward to the ISCO, the disk extends until it terminates, as stable circular motion is no longer possible inside the ISCO and particles generally plunge in the black hole. Therefore, the inner edge of the accretion disk is usually taken to be at the ISCO. In our simplified model, the black hole jet is assumed to originate near the warp radius, and consequently, its direction undergoes precession relative to the black hole spin axis. Furthermore, we require that the jet’s precession rate matches that of a particle moving on a spherical orbit at the warp radius. In a simplified picture, our model can be envisioned as a rigid, tilted thin disk with a slender rod rigidly attached perpendicular to the disk surface. The rod then precesses around the vertical axis. Although this model is indeed idealized for describing a realistic black hole accretion disk-jet system, it captures the essential kinematic features and is expected to yield physically meaningful results. In fact, parameter constraints derived from this model for Kerr black holes have been found to be comparable to those obtained from certain numerical simulations~\cite{Wei,Cui2}. Based on this agreement, we proceed to examine whether the central supermassive object in M87\textsuperscript{*} could be a KTN black hole, and use the observed jet precession to constrain its parameters. The analysis also includes the previously unexplored case of naked singularities. To clarify the foundations of our modeling approach, we summarize the key assumptions below~\cite{Wei}: (a) The motion of particles at different radii within the thin disk can be accurately described by the spherical orbits with constant tilt angle; (b) The jet is assumed to originate near the warp radius and is perpendicular to the local disk surface; (c) The precession axis coincides with the black hole spin axis. It is worth emphasizing that, unlike previous studies where the disk is tilted with respect to the equatorial plane, the tilted disk in the KTN spacetime is inclined relative to a non-equatorial reference plane due to the presence of the NUT charge.

In order to constrain the black hole parameters using the observed jet precession, we need to calculate the precession angular velocity of the spherical orbits for each set of parameters in the parameter space. In the previous section, we provided the method for calculating this angular velocity in natural units; in this section, we restore it to SI units and present the final expression
\begin{equation}\label{eqOmegasi}
    \Omega=\omega_{t}\frac{M_{\odot}}{M}\left(\frac{c^3}{GM_{\odot}}\right)\approx 6.40982\times10^{12}\times\omega_{t}\frac{M_{\odot}}{M}\,\left(\text{year}^{-1}\right),
\end{equation}
where $M_{\odot}$ denotes the solar mass, and $M$ represents the mass of the M87* black hole~\cite{Akiyama1}. For clarity, the precession angular velocity $\omega_t$ can be written in the functional form
\begin{equation}
\omega_t = f\left(r_{\text{w}}, a, n, E(r_{\text{w}}, a, n, \theta_0, \zeta), L(r_{\text{w}}, a, n, \theta_0, \zeta)\right).
\end{equation}
Here $r_{\text{w}}$ is the warp radius, which must satisfy $r_{\text{w}} \geq r_{\text{ISSO}}$. According to the observational data of the M87* jet, the tilt angle is given by $\zeta = 1.25^\circ \pm 0.18^\circ$. For simplicity, we neglect the small uncertainty in $\zeta$ and fix it at $1.25^\circ$.

Based on the derivations in the previous sections, we outline the procedure for constraining the black hole parameters as follows. We consider the parameter space $\{a, n, r_{\text{w}}\}$ and use interpolation to determine the warp radius $r_{\text{w}}$ corresponding to $\Omega = 0.56$ for each $\{a, n\}$ pair. This is equivalent to solving the constraint equation
\begin{equation}
\Omega(a, n, r_{\text{w}}) = 0.56,
\end{equation}
which defines a constraint surface in the parameter space.
Note that although the observed jet precession angular velocity is $0.56 \pm 0.02$ rad/yr~\cite{Cui}, we fix it at $0.56$ to reduce computational cost.

To compute $\Omega$ for a given set of parameters, we first solve Eq.~\eqref{eqtheta0} to obtain the angular position $\theta_0$ of the circular orbit, and substitute it into Eqs.~\eqref{eqHESO} and \eqref{eqHLSO} to calculate the energy $E$ and angular momentum $L$ of spherical orbit. We then solve Eqs.~\eqref{eqtdotmotionSO}-\eqref{eqpthetadotmotionSO} and apply Eq.~\eqref{eqomegat} to obtain the precession angular velocity $\omega_t$, which is finally used in Eq.~\eqref{eqOmegasi} to obtain the corresponding $\Omega$ for each $\{a, n, r_{\text{w}}\}$ set.

\subsection{Constraints from the jet precession of M87*}
\begin{figure}[!htbp]
    \centering{
			\subfigure[]{\includegraphics[width=6.7cm]{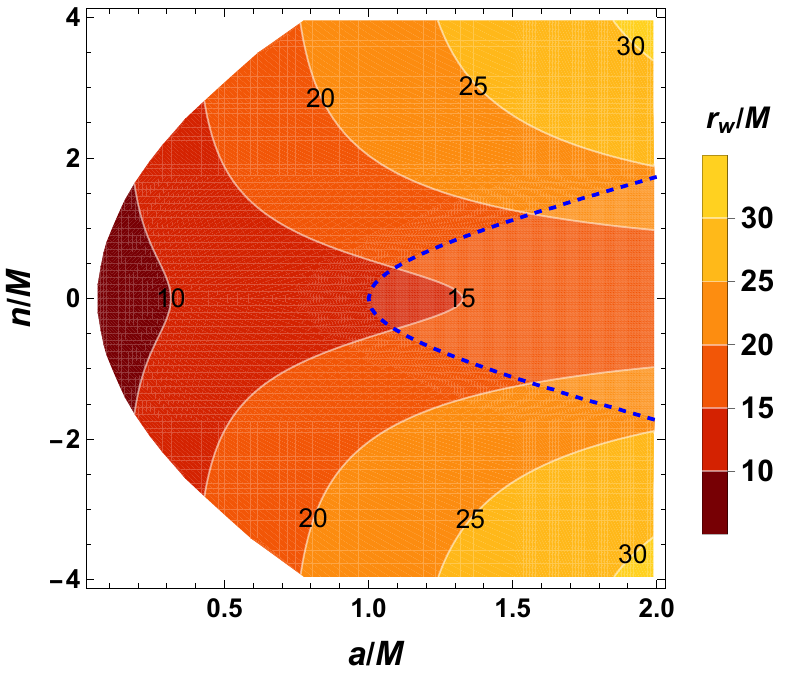}\label{figprwan11a}}
            \subfigure[]{\includegraphics[width=7cm]{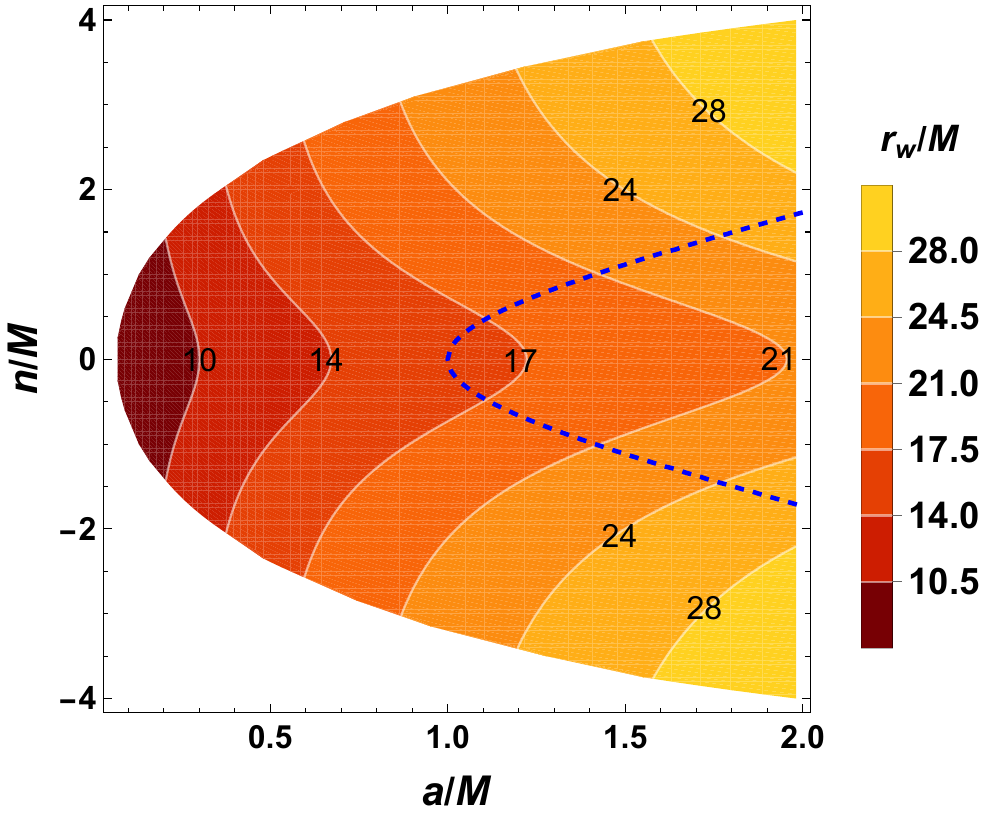}\label{fignrwan11b}}
			\caption{Variation of the warp radius $r_{\text{w}}$ with respect to the black hole spin $a$ and NUT charge $n$. The blank regions in the figure represent parameter spaces excluded by jet precession, where $r_{\text{w}} < r_{\text{ISSO}}$ and thus violate the physical condition. The arched regions enclosed by blue dashed lines correspond to the cases of naked singularities. (a) Prograde accretion disk; (b) Retrograde accretion disk.
}\label{figrwan11}
    }
\end{figure}

Based on the computational procedure described above, in Fig.~\ref{figrwan11}, we show the variation of the warp radius $r_{\text{w}}$ of the tilted accretion disk as a function of the black hole spin $a$ and the NUT charge $n$, calculated from the observed jet precession angular velocity of $0.56$. The region with small spin and large NUT charge is excluded by the physical condition $r_{\text{w}} \geq r_{\text{ISSO}}$. Moreover, the excluded parameter region is larger for retrograde disks, and even in the RS-KTN case, part of the parameter space is ruled out. We also observe that the black hole spin must be nonzero. As pointed out in Sec.~\ref{subsecTNSO}, the spherical orbits in the Taub-NUT spacetime do not exhibit LT precession. Therefore, M87* cannot be a purely NUT-charged object without spin. It is evident that the case of naked singularities is not excluded, and connects continuously to the KTN black hole regime. From an observational perspective, this implies that the possibility of a naked singularity cannot be definitively ruled out. In addition, we find that the sign of the NUT charge cannot be distinguished, since the warp radius is symmetric about $n = 0$. As the absolute value of $n$ increases, the warp radius also increases. This is because the precession angular velocity $\omega_{t}$ increases with $|n|$, but decreases with orbital radius $r$. As a result, for a fixed precession rate, a larger $|n|$ leads to a larger $r$.

\begin{figure}[!htbp]
    \centering{
			\subfigure{\includegraphics[width=7cm]{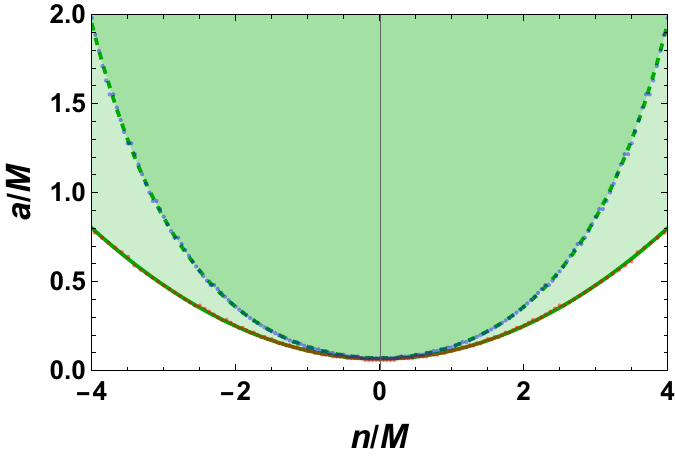}}
			\caption{Constraints on the black hole parameter space $(a, n)$ for prograde and retrograde accretion disks, corresponding to Fig.~\ref{figrwan11}. The blank region represents the parameter space excluded by both prograde and retrograde disk configurations. The area composed of both dark green and light green indicates the region allowed for prograde disks. Red dots along the boundary originate from the data points on the transition curve between the excluded (blank) and allowed (colored) regions in Fig.~\ref{figprwan11a}, and the green solid curve is a fitting result for these points. The dark green region represents the parameter space allowed for retrograde disks, with blue dots taken from the boundary between blank and colored regions in Fig.~\ref{fignrwan11b}. The green dashed curve shows the fit to these data points.
}\label{figfitConstrain12}
    }
\end{figure}

Through the physical condition $r_{\text{w}} \geq r_{\text{ISSO}}$, we have excluded a portion of the parameter space, corresponding to the blank regions shown in Fig.~\ref{figrwan11}. To quantitatively describe these excluded regions, we perform a fitting on the data located along the boundary between the blank and colored areas in the figure, and obtain a relation
\begin{equation}\label{eqanconstraint}
    a\ge \begin{cases}
	0.0459706\,n^2+0.0662973&		,\ \text{for\ prograde\ disk,}\\
	0.841685\,e^{0.0733814\,n^2}-0.771071&		,\ \text{for retrograde disk,}\\
\end{cases}
\end{equation}
characterizing the allowed region in the $(a, n)$ parameter space. It is worth noting that our scan is limited to $a \in [0, 2M]$ and $n \in [-4M, 4M]$, which is sufficient for the current study. The fitting results are shown in Fig.~\ref{figfitConstrain12}, where the data align almost perfectly with the fitted curve, demonstrating excellent agreement.

Therefore, by using the observed precession rate of the M87* jet, we have constrained the black hole parameter space, excluding regions with low spin and large NUT charge. Importantly, the constraints differ between the prograde and retrograde cases, as the jet precession observation does not provide information about the direction of precession. For this reason, a parallel discussion of prograde and retrograde disk configurations is carried throughout this work.

\subsection{The size of inner disk}
After studying the warp radius of tilted accretion disk, we proceed to investigate the inner part of the disk, referred to as the ``inner disk". It is defined as the region between the ISCO and the warp radius $r_{\text{w}}$. The inner disk typically lies in the equatorial plane in Kerr spacetime \cite{BardeenPetterson}. However, in the presence of NUT charge, the inner disk is no longer confined to the equatorial plane but is composed of a set of non-equatorial circular orbits. These orbits are parallel to the equator but have angular positions significantly deviating from $\pi/2$, as studied in detail in Sec.~\ref{secKNTCO}. Nevertheless, the size of non-tilted inner disk, composed of non-equatorial circular orbits, can still be naturally defined as
\begin{equation}
    R = r_{\text{w}} - r_{\text{ISCO}},
\end{equation}
where $r_{\text{ISCO}}$ has been evaluated in Sec.~\ref{subsecISCO} following the methods presented there.

\begin{figure}[!htbp]
    \centering{
			\subfigure[]{\includegraphics[width=6.8cm]{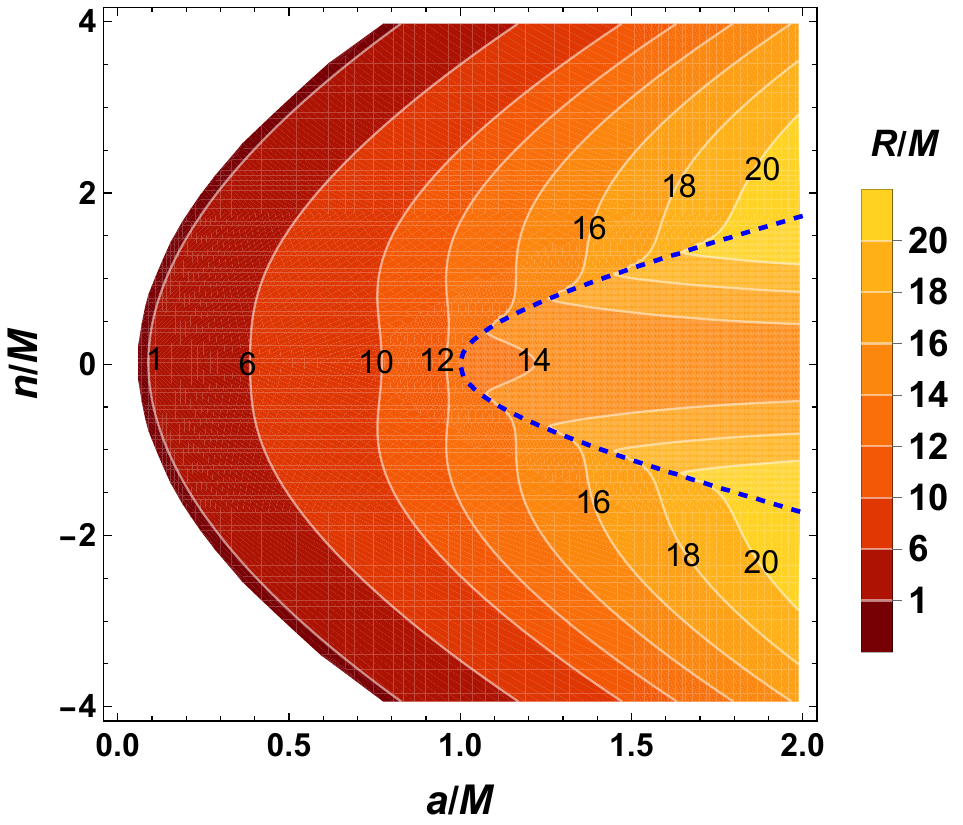}\label{figpRsize13a}}
            \subfigure[]{\includegraphics[width=7cm]{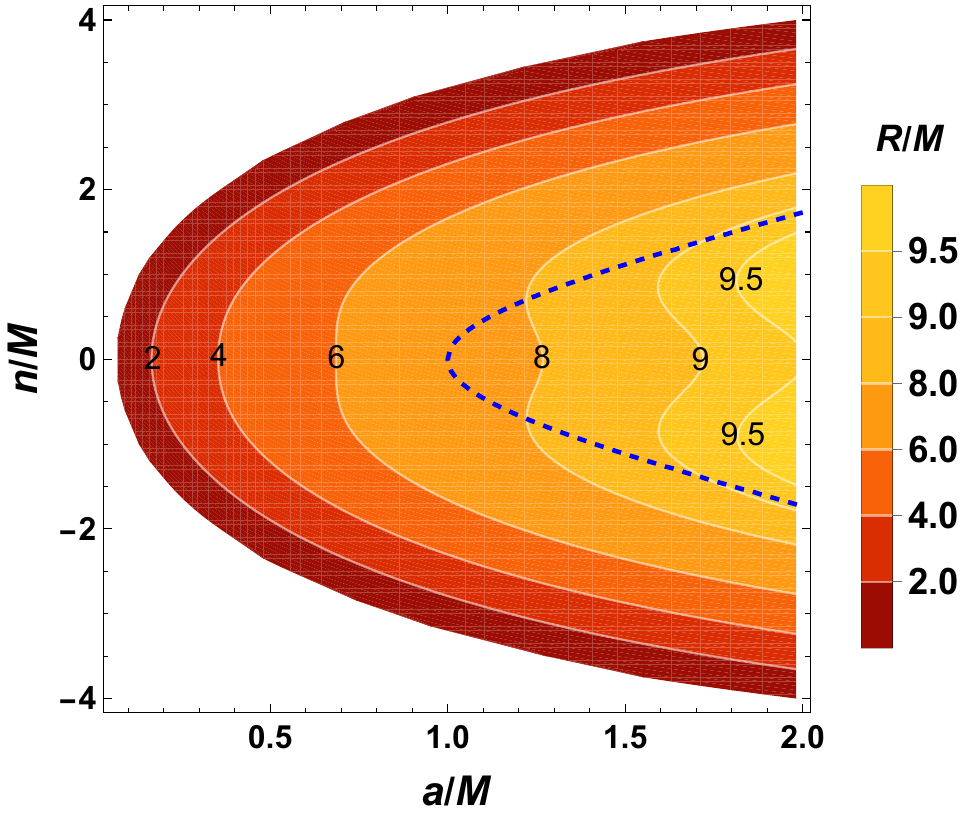}\label{fignRsize13b}}
			\caption{Variation of the inner disk size $R$ with respect to the black hole spin $a$ and NUT charge $n$. The blank regions denote the excluded parameter space, consistent with the exclusions shown in Fig.~\ref{figrwan11}. The arched region enclosed by the blue dashed line corresponds to the naked singularity regime. (a) Prograde disk. (b) Retrograde disk.
}\label{figRsize13}
    }
\end{figure}

In Fig.~\ref{figRsize13}, we present the variation of the inner disk size $R$ with respect to the black hole spin $a$ and NUT charge $n$. The results show that the size $R$ remains independent of the sign of $n$, with the plots being symmetric about $n = 0$. For the case of KTN black holes, the inner disk size decreases with increasing $|n|$, regardless of whether the disk is in a prograde or retrograde configuration. However, in the case of naked singularities, a distinct behavior emerges: for prograde disks, $R$ increases with $|n|$, while for retrograde disks, $R$ first increases and then decreases, showing a non-monotonic dependence on $|n|$. These results indicate that although the warp radius varies continuously between black hole and naked singularity cases, the inner disk structure is significantly affected in the naked singularity regime. This difference may be attributed to the unique properties of the ISCO in the presence of a naked singularity.

\section{Discussions and conclusions}\label{secConclusions}

In this work, we studied the precession of the spherical orbits in the KTN spacetime and explored their astrophysical implications. The presence of the NUT charge breaks the $\mathbb{Z}_2$ symmetry of the spacetime, and it reflects in the geodesic structure. Consequently, circular orbits lie on the non-equatorial planes. As a result, the spherical orbits, which generalize circular ones, also exhibit a breakdown of the reflection symmetry in the $\theta$ direction.

In our analysis of the spherical orbits, we initially explored the corresponding circular orbits, determining their energy, angular momentum, and angular position. Utilizing the Hamiltonian framework, we derived the equations of motion and identified the associated conserved quantities. Our investigation revealed that while the sign of the NUT charge $n$ does not alter the energy and angular momentum values, their values increase with the absolute value of $n$. The angular position $\theta_0$ of these circular orbits is contingent on the sign of $nL$: specifically, $\theta_0 > \pi/2$ for $nL > 0$ and $\theta_0 < \pi/2$ for $nL < 0$. Circular orbits corresponding to opposing values of $n$ exhibit symmetry about the equatorial plane, with their departure from the equator expanding with $|n|$. The impact of the black hole spin $a$ on this deviation varies for prograde and retrograde orbits: in the prograde scenario, increasing $a$ augments the deviation from $\pi/2$, while the retrograde situation displays an inverse relationship. Furthermore, we calculated the position of the ISCOs. As the absolute value of $n$ increases, the ISCO's radius increases, distancing itself from the equator. Remarkably, for prograde ISCOs, there are sharp corners in the radial and angular positions near the extremal black hole within the parameter region.

Building upon the analysis of circular orbits, we investigated the characteristics and precession of the spherical orbits to model particle motion within tilted accretion disks. We formulated analytical expressions for their energy and angular momentum and introduced a deviation angle $\zeta$ to quantify the difference between the upper turning point $\theta_{\min}$ and the angular position $\theta_0$ of a circular orbit at the same radius. Subsequently, we computed the energy and angular momentum of the spherical orbit. While the NUT charge has a minor impact on the energy and angular momentum of orbits inside the ISSO, our focus shifted to those beyond the ISSO, where these quantities are nearly invariant to the sign of $n$. Additionally, we determined the radius $r_{\text{ISSO}}$ of the ISSO, serving as a lower bound for the warp radius $r_{\text{w}}$. Moreover, we demonstrated that solving for the ISCO or ISSO using the extreme points of energy or angular momentum is equivalent to the conventional effective potential method. Our exploration extended to analyzing the precession of the spherical orbits. In the Taub-NUT spacetime with a zero spin, the spherical orbits transition into tilted circular orbits without precession. Introducing a nonzero spin was essential for investigating the impact of the NUT charge on precession. By numerically solving the complete equations of motion, we illustrated how the NUT charge disrupts symmetry under simultaneous space-time reflection about the circular orbit in $\theta$-motion, resulting in a slight discrepancy between the deviation angle and the tilt angle. However, this discrepancy was negligible enough to consider $\zeta$ as the tilt angle in practical observational scenarios. We then derived the precession angular velocity from the trajectory and unveiled that this velocity remains indifferent to the sign of the NUT charge, yet it increases with the absolute value of the charge.

In the final part of the paper, we applied the spherical orbits to model the motion of particles near the warp radius in the tilted accretion disk around the M87 black hole described by the KTN metric. Under certain assumptions, the precession of these orbits is synchronized with the jet precession. Using the observed value of the jet precession angular velocity, we constrained the black hole parameter space, which included both black holes and naked singularities. Since the direction of jet precession is not determined observationally, we considered both prograde and retrograde disks. The results indicated that the sign of the NUT charge cannot be distinguished. Regions with small spin $a$ and large $n$ were excluded, and in the retrograde case, the excluded region was larger, extending to high spin and large NUT charge. The naked singularity region remained allowed, suggesting that jet precession alone cannot distinguish black holes from naked singularities. We also fitted the boundary of the allowed region and provided the explicit relation in Eq.~\eqref{eqanconstraint}. In addition to parameter constraints, we investigated how black hole parameters affect the inner region of the accretion disk. For both prograde and retrograde disks in the black hole regime, the size $R$ of inner disk  decreases with $|n|$. However, in the naked singularity regime, a different behavior emerged: for prograde disks, $R$ increases with $|n|$, whereas for retrograde disks, $R$ increases and then decreases, showing a non-monotonic dependence on $|n|$.

Overall, we explored the distinctive features of the spherical orbits in the KTN spacetime, constrained black hole parameters using jet precession observations, and examined their impact on the structure of accretion disk. Our results suggest that a nonzero NUT charge remains a viable possibility and that the jet precession cannot currently distinguish between black holes and naked singularities. This highlights the potential of jet precession as a tool to constrain the parameters of black holes and test the theory of gravity. In the future, combined with other observations, some parameters can be constrained more accurately.

\section*{Acknowledgements}
	This work was supported by the National Natural Science Foundation of China (Grants No. 12475055, and No. 12247101), the Basic Research Foundation of Central Universities (Grant No. lzujbky-2024-jdzx06), the Natural Science Foundation of Gansu Province (Grant No. 22JR5RA389), and the `111 Center' under Grant No. B20063.
	
\bibliographystyle{unsrt}

\end{document}